\documentclass[aps, twocolumn, prl, showpacs,amsmath,amssymb,superscriptaddress]{revtex4-1}
\usepackage{amsmath,amsfonts,amssymb,graphics,graphicx,epsfig,color,times,indentfirst,layout,hyperref,subfigure,multirow,bm}
\usepackage{hyperref,soul}
\usepackage{ulem}
\hypersetup{linktocpage,,colorlinks=true}
\usepackage{threeparttable}

\def\beq{\begin{equation}}
\def\eeq{\end{equation}}
\def\bea{\begin{eqnarray}}
\def\eea{\end{eqnarray}}

\begin{document}

\title{Evolution of entanglement spectra under generic quantum dynamics}

\author{Po-Yao Chang$^{1,2,3}$, Xiao Chen$^4$, Sarang Gopalakrishnan$^5$, and J. H. Pixley}
\affiliation{Department of Physics and Astronomy, Center for Materials Theory, Rutgers University, Piscataway, NJ 08854 USA \\
$^2$ Max Planck Institute for the Physics of Complex Systems, N\"{o}thnitzer Strasse 38, D-01187 Dresden, Germany \\
$^3$ Department of Physics, National Tsing Hua University, Hsinchu 30013, Taiwan\\
$^4$ Kavli Institute for Theoretical Physics, University of California at Santa Barbara, Santa Barbara, CA 93106, USA \\
$^5$ Department of Physics and Astronomy, CUNY College of Staten Island, Staten Island, NY 10314, USA, and Physics Program and Initiative for the Theoretical Sciences, CUNY Graduate Center, New York, NY 10016, USA}

\begin{abstract}
%{\clb

We characterize the early stages of the approach to equilibrium in isolated quantum systems through the evolution of the entanglement spectrum. We find that the entanglement spectrum of a subsystem evolves with at least three distinct timescales. First, on an $o(1)$ timescale, independent of system or subsystem size and the details of the dynamics, the entanglement spectrum develops nearest-neighbor level repulsion. The second timescale sets in when the light-cone has traversed the subsystem. Between these two times, the density of states of the reduced density matrix takes a universal, scale-free $1/f$ form; thus, random-matrix theory captures the \emph{local} statistics of the entanglement spectrum but not its global structure. The third time scale is that on which the entanglement saturates; this occurs well after the light-cone traverses the subsystem. Between the second and third times, the entanglement spectrum \emph{compresses} to its thermal Marchenko-Pastur form. 
These features hold for chaotic Hamiltonian and Floquet dynamics as well as a range of quantum circuit models.
\end{abstract}

\maketitle

Understanding how an isolated quantum system reaches thermal equilibrium is a central problem in quantum statistical physics.
Substantial progress has been made on the late-time aspects of thermalization, based on the eigenstate thermalization hypothesis~\cite{deutsch_eth, srednicki_eth, rigol2008thermalization, Cardy2014, Garrison2018}, which implies that small enough subsystems are well described by thermal density matrices if one waits long enough for information to have traversed the entire system. 
Much numerical \cite{rigol2008thermalization,Rigol-2012,Garrison2018} and experimental \cite{Kaufman2016} evidence now exists for eigenstate thermalization.
However, the mechanism  
by which a local density matrix goes from being disentangled to being fully thermal---the \emph{process} of thermalization---is still poorly understood. 
Some coarse grained features of the thermalization process have recently been characterized numerically, as well as through the study of random unitary circuits (RUCs)~\cite{Keyserlingk2017, Nahum2017,nrh,Khemani2017, rpv, pai2018,Chan2017,chen2018quantum}. In special limits of RUCs (namely, the limit of large on-site Hilbert space or Clifford circuits), and fine-tuned models such as the self-dual kicked Ising model~\cite{bertini}, exact solutions are available for entanglement growth and the scrambling of local operators. These solvable cases, however, are non-generic, and miss important aspects of the generic  thermalization process.

The present work addresses the dynamics of entanglement and thermalization at early times in  \emph{generic} systems (i.e.~nonintegrable models with a low-dimensional on-site Hilbert space): here, 
the entanglement spectrum [i.e., the eigenvalues of the reduced density matrix (RDM)]~\cite{znidaric2012, hamma1, hamma2, hamma3, Mierzejewski2013} evolves in a highly nontrivial way that is not even qualitatively captured by the entanglement entropy. This behavior is absent in the aforementioned exactly solvable limits. 
The picture that emerges is 
independent of how the dynamics is generated,
holding for Hamiltonian, Floquet, and temporally random dynamics; for systems with and without conservation laws; and for chaotic as well as many-body localized systems. Here, we focus on Hamiltonian dynamics and RUCs; for other cases see~\cite{suppmat}. 
%We have demonstrated this universality for an array of models but for simplicity we only present a few representative cases in the main text; the other models are detailed in~\cite{suppmat}.
%

\begin{figure}[!bt]
\begin{center}
\includegraphics[width = 0.48\textwidth]{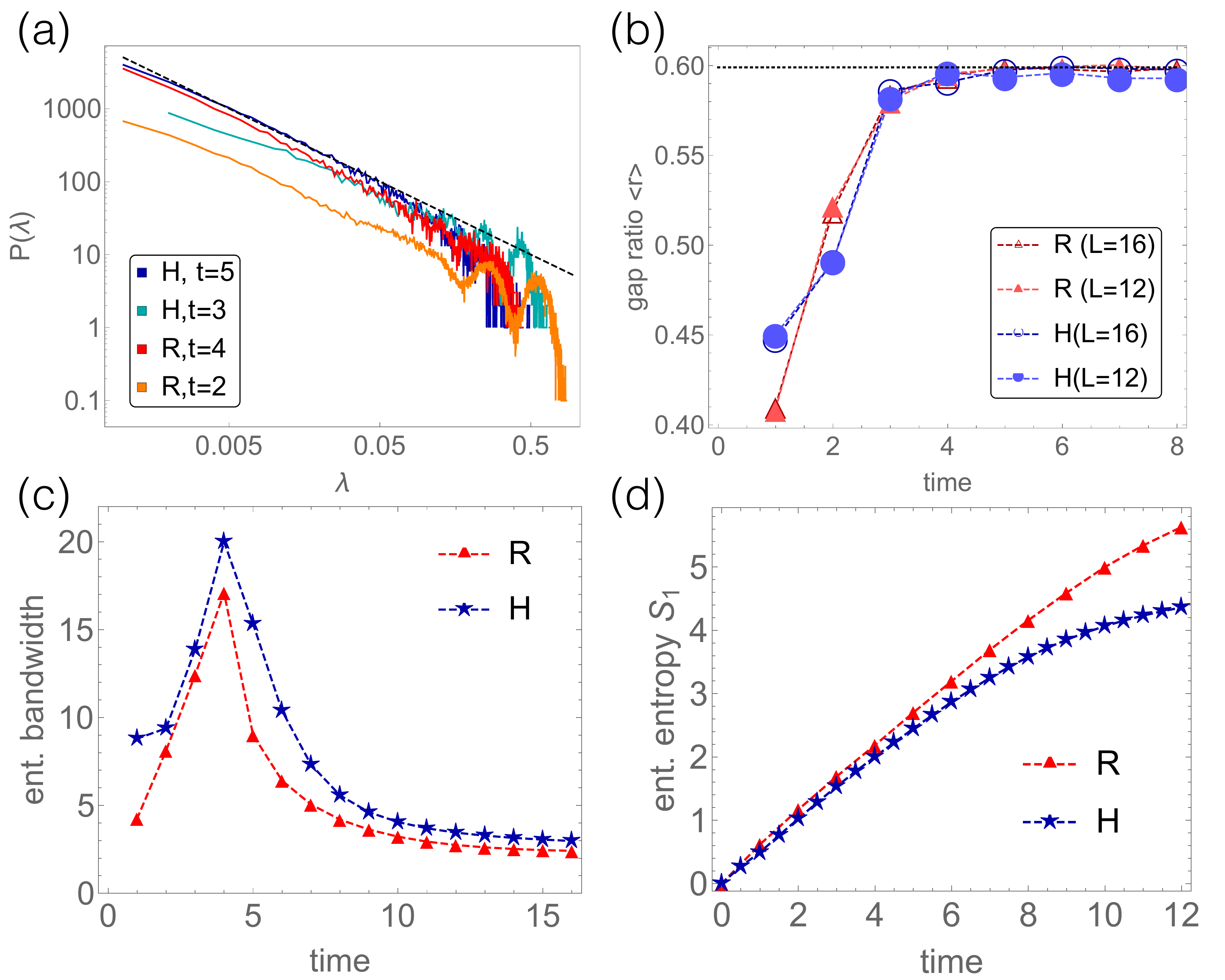}
\caption{\textit{Spectral properties of the reduced density matrix under generic time evolution.} (a)~Spectral density of the reduced density matrix for random unitary (R) and chaotic Hamiltonian (H) dynamics at early times; this follows a $1/f$ distribution (dashed line). (b)~Adjacent gap ratio of the entanglement spectrum, as a function of time, comparing R and H dynamics. Colors denote the model (red for R and blue for H); for each color, empty symbols are for system size $L = 12$ and subsystem size $l_A = 6$, whereas filled symbols are for $L = 16, l_A = 8$. 
The random-matrix prediction [$\langle r\rangle\approx 0.599$] is marked with a dashed black line. 
(c, d)~Evolution of entanglement bandwidth $w$ and von Neumann entanglement entropy $S_1$, for R and H dynamics with $L = 16, l_A = 8$. For both R and H dynamics the entanglement bandwidth grows until $t = L/2$ (for RUC's this growth is linear), then shrinks, whereas the entanglement entropy keeps growing. The saturation value of $S_1$ is higher for R dynamics than H dynamics, due to the absence of conservation laws.}
\label{fig1}
\end{center}
\end{figure}

In this work, we show that the process of thermalization takes place in at least three stages; our main new results are that the entanglement spectra behave universally even at relatively early times as demonstrated in Fig.~\ref{fig1}, although its early- and late-time properties belong to different universality classes.
To explain these regimes, we introduce multiple characteristic timescales in the entanglement evolution: (i) the timescale on which the entanglement spectrum develops nearest-neighbor level repulsion; (ii) the timescale on which the rank of the density matrix [i.e., the R\'enyi entropy $S_0$ in Eq.~\eqref{eqn:Sa}] saturates; and (iii)~the timescale on which the reduced density matrix saturates to its late-time behavior. 
One of our main results is that timescale~(i) is independent of system and subsystem size, and largely insensitive to the nature of the dynamics.
A second main result is that the spectrum of the reduced density matrix between timescales~(i) and (ii) exhibits a universal, scale-invariant $1/f$ density of states. This distribution spreads over increasingly many decades as time passes, until we hit timescale~(ii). Once again, this behavior is present in all the models we have considered, but is absent in the exactly solvable limits. 
%
%In the solvable models, the entanglement spectrum has a trivial shape as all eigenvalues of the RDM coincide; only its rank evolves.
%
Finally, between timescales (ii) and (iii) the range of the $1/f$ distribution shrinks, and narrows toward the late-time Marchenko-Pastur form~\cite{hamma1}; during this entire process the entanglement entropy is still growing. For quantum circuits, which have a strict light cone, there is a sharp transition between these regimes, set by the subsystem size. For Hamiltonian dynamics this is rounded into a crossover (due to the exponential tails in the Lieb-Robinson bound~\cite{Lieb-1972}) but the two temporal regimes are still clearly distinguished in practice (Fig.~\ref{fig1}). 
Both of our main findings are absent in exactly solvable limits, where  the entanglement density of states is a delta function at all times, and consequently the nearest-neighbor level spacing is not defined. 
%

%For the quantum chaotic models under consideration 
We capture level statistics beyond nearest-neighbor using an appropriate \textit{entanglement} spectral form factor. At short times the spectral form factor of the entanglement spectrum has a ``ramp'' feature characteristic of level repulsion, but does not quantitatively behave as random-matrix theory would predict. Further, the spectral form factor drifts with time until very late times when the entanglement has saturated; only then does it take on its universal shape dictated by random matrix theory.
Thus our results clarify the sense in which such systems are ``locally thermal'': although the coarse structure of the reduced density matrix is far from that of a thermal state, its ``short-distance'' level statistics  look thermal.

\begin{figure}[tb]
\begin{center}
\includegraphics[width=\linewidth]{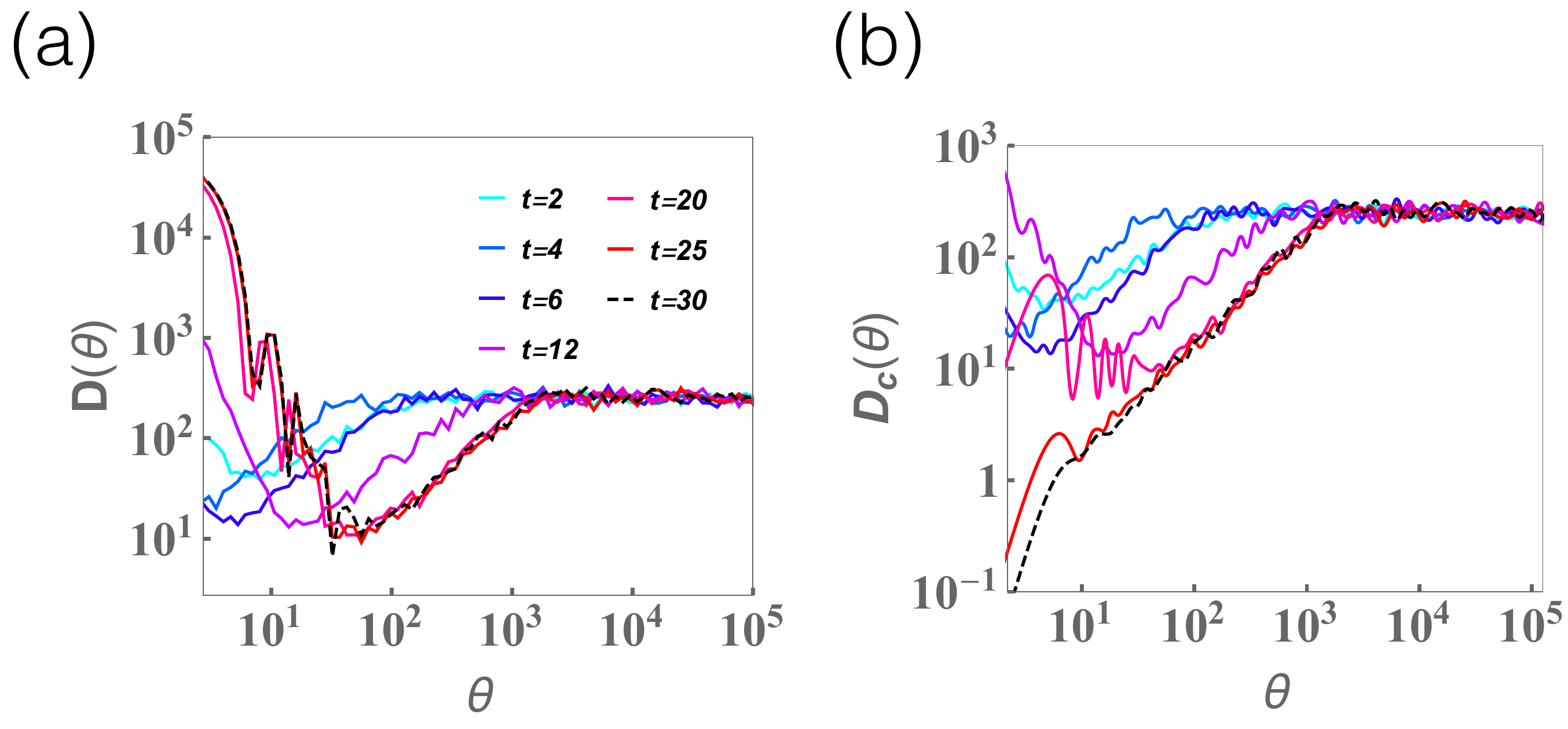}
\caption{{\it The
entanglement  spectral form factor $D(\theta)$ as a function of $\theta$ for various different times in the RUC without a conservation law.} (a) The disconnected ESFF $D(\theta)$ and (b) the connected ESFF $D_c(\theta)$ at various times. The legend in (a) is shared across these figures.
}
\label{formfactors}
\end{center}
\end{figure}

\textit{Models}.---The main results outlined above were checked for a variety of models, both under discrete-time evolution (i.e., quantum circuits) and continuous-time Hamiltonian evolution. The quantum circuits considered here all involve time-evolution operators of the form
$U(t) = U(t,t-1)U(t-1,t-2) \cdots U(1,0)$, where
\begin{align}
U(t',t'\!-\!1)\!=\!&\bigotimes_{i \in 2 \mathbb{Z}}U_{i,i+1}\! (t',t'\!-\!1) \!\!\!\bigotimes_{i \in 2 \mathbb{Z}+1}\!\!\!\!U_{i,i+1}\! (t',t'\!-\!1),
\end{align}
with $i$ being the site index and $U_{i,i+1}$ being unitary matrices. When written as a matrix in the many-body Hilbert space the gates are very sparse, and therefore we simulate them exactly using sparse matrix-vector multiplication.
 In the main text we present results for circuits in which these unitaries are randomly chosen at each point in space and time; we draw them either completely randomly (with Haar measure) or from an ensemble of random matrices with a single conservation law~\cite{Khemani2017}. 
We have also simulated the Floquet versions of these circuits, but find no noticeable differences in entanglement spectra between the temporally random and Floquet cases.
One other case---a Floquet model that is many-body localized~\cite{Chan2018PRL} rather than chaotic---is shown in~\cite{suppmat}. Although the evolution of $S_1$ is very different in this case, the entanglement spectrum still shows level repulsion and a $1/f$ distribution in its bulk: the absence of chaos only manifests itself in the properties of the largest few Schmidt coefficients (lowest entanglement energies). 
%In~\cite{suppmat} we consider a Floquet model that is many-body localized rather than chaotic~\cite{Chan2018PRL}.

%We use exact numerical time evolution to simulate the models considered here.

To study Hamiltonian evolution we consider the Ising model with both transverse and longitudinal fields: 

\beq
H = \sum_i J \sigma^z_{i}\sigma^z_{i+1}+h_x \sigma^x_i +h_z \sigma^z_i
\label{eqn:HI}
\eeq
where $\sigma^\alpha_i$ are spin-$1/2$ Pauli operators. For our simulations we choose the parameters $(h_x/J, h_z/J) = (0.9045, 0.809)$, corresponding to a nonintegrable regime in which thermalization is known to be fast~\cite{Kim-2013,Kim-2015} (in what follows we set $J=1$ as the unit of energy for Hamiltonian dynamics). We use a Krylov-space method to efficiently time-evolve the state~\cite{brenes2019}.

\textit{Measured quantities}.---The RDM of any subsystem has non-negative real eigenvalues $\{ \lambda_n \}$. Since broad distributions are present, it is helpful to work with the entanglement spectrum, which has eigenvalues $\{ E_n \} = \{ -\log \lambda_n \}$. The entanglement density of states is a probability density over these eigenvalues [given by $\varrho_S(E) = D^{-1}\sum_n\delta(E-E_n)$ where $D$ is Hilbert space dimension of subsystem $A$], and the entanglement bandwidth is the width of this probability distribution~\footnote{Because very small eigenvalues of $\rho$ are contaminated by machine precision, we define the width as twice the distance from the median to the 25th percentile of the entanglement spectrum (i.e., median to 75th percentile of the spectrum of the RDM). This matches the interquartile range when both can be reliably computed.}. The Renyi entropies are moments of the $\{ \lambda_n \}$:
\beq
S_\alpha \equiv \frac{1}{1-\alpha} \log\left( \sum\nolimits_n \lambda_n^{\alpha} \right).
\label{eqn:Sa}
\eeq
There are three special limits: as $\alpha \rightarrow 0$, $S_\alpha$ returns the rank of the reduced density matrix; as $\alpha \rightarrow \infty$, $S_\alpha$ picks out the largest eigenvalue of the reduced density matrix; and as $\alpha \rightarrow 1$, $S_\alpha$ approaches the Von Neumann entanglement entropy. 

We quantify level statistics via the adjacent gap ratio $r$~\cite{oganesyan_huse},
\begin{align}
r_m \equiv \frac{{\rm min}(\delta_m,\delta_{m+1})}{{\rm max}(\delta_m,\delta_{m+1})},
\end{align}
where $\delta_m = E_m - E_{m-1}$ and the $E_m$ are arranged in ascending order.
 The average adjacent gap ratio takes the value $\langle r \rangle \approx 0.599$ for the Gaussian unitary ensemble (GUE); its probability distribution also approaches a universal form~\cite{oganesyan_huse}.
% We compute the probability distribution of $r_m$, as well as its average over the entire spectrum and over specific \cite{suppmat}. 

The adjacent gap ratio is only sensitive to the level repulsion of neighboring eigenvalues. To quantify ``longer-range'' level repulsion we  study the spectral form factor of the entanglement spectrum, which is the Fourier transform of the correlation function between two levels in the entanglement spectrum.
We term this the ``entanglement spectral form factor'' (ESFF).  
The ESFF  characterizes  the {\it global} level statistics of the entanglement spectrum, and is expressed as:
\beq
D(\theta) \equiv \left\langle \sum\nolimits_{n,m} e^{i \theta(E_n - E_m)} \right\rangle.
\eeq
Here $\theta$ denotes an auxiliary ``time'' variable that is conjugate to the entanglement ``energy.'' For a GUE random matrix, 
the spectral form factor has
a linear growth in $\theta$, called the {\it ramp}, followed by a sudden saturation, reaching its {\it plateau} value~\cite{GUHR1998}.
A precise {\it ramp-plateau} structure can be obtained by subtracting out the disconnected parts $\vert\langle \sum\nolimits_{n} \exp(i \theta E_n )\rangle \vert^2$,
which defines the connected ESFF $D_c(\theta) = D(\theta) - \vert\langle \sum\nolimits_{n} \exp(i \theta E_n )\rangle \vert^2$.
These form factors have the advantage of capturing gap correlations beyond nearest neighbor, but the disadvantage of being sensitive to the overall entanglement density of states (DOS), which as we have seen in Fig.~\ref{fig1} (c) are strong. Note that the ESFF is not the unique spectral form factor one can construct for the reduced density matrix; we could instead have constructed a spectral form factor from the eigenvalues of the reduced density matrix~\cite{suppmat}. However, the ESFF has the crucial advantage that its asymptotic large-$\theta$ behavior is set by the \emph{large} Schmidt coefficients, and is therefore sensitive to the late stages of the thermalization process.

Under Hamiltonian dynamics, the eigenstate thermalization hypothesis implies that at late times the reduced density matrix takes the form $\rho_A = \exp(-H_A/T)$, where $T$ is the temperature set by the global energy density~\cite{Garrison2018}. 
Thus, the ESFF matches the spectral form factor of the Hamiltonian (projected into the subsystem), up to rescaling. 
 On the other hand, under random unitary dynamics, 
even when there is a conservation law, the conserved quantity is not the generator of the dynamics.
Hence, the ESFF acts as a measure of how random the state is, and its late-time structure is what one would predict from a random pure state~\cite{Chen2017}. 
We find that both spectral form factors settle down to a time-independent function that is consistent with the shape predicted from random matrix theory, once the entanglement entropy has completely saturated (see Fig.~\ref{formfactors} and~\cite{suppmat}).

\textit{Purely random circuits}.---We first discuss our results for the purely random case. In this case each gate is picked Haar-randomly and independently at each space and time point. We have already outlined the main results in Fig.~\ref{fig1} and now discuss them in greater detail.
The distribution of RDM eigenvalues becomes broad at short times ($t< l_A/2$, where $l_A$ is the size of the subsystem) following a universal scale free $1/f$ distribution [Fig.~\ref{fig1} (a)]. The entanglement level statistics rapidly approaches its random-matrix value on an $o(1)$ timescale [Fig.~\ref{fig1} (b)], that is independent of the system and subsystem size.  The entanglement bandwidth initially grows linearly in time, out to a time $t = l_A/2$ when the light cone hits the edge of the subsystem and then decays algebraically to a small steady state value [Fig.~\ref{fig1} (c)]. During this short time dynamical process 
 the entanglement entropy continues to grow until it saturates at time scale set by the system size [Fig.~\ref{fig1} (d)]. 
In Fig.~\ref{formfactors} we show the behavior of the ESFF 
in this model, for $L = 20, l_A = 8$. The ESFF develops a ramp-plateau structure at early times, corresponding to the short timescale on which
level repulsion sets in among the entanglement ``energy levels". However, the overall shape of the ESFF drifts over time, until the entanglement bandwidth and entanglement entropy have saturated.

\begin{figure}[tb]
\begin{center}
\includegraphics[width=\linewidth]{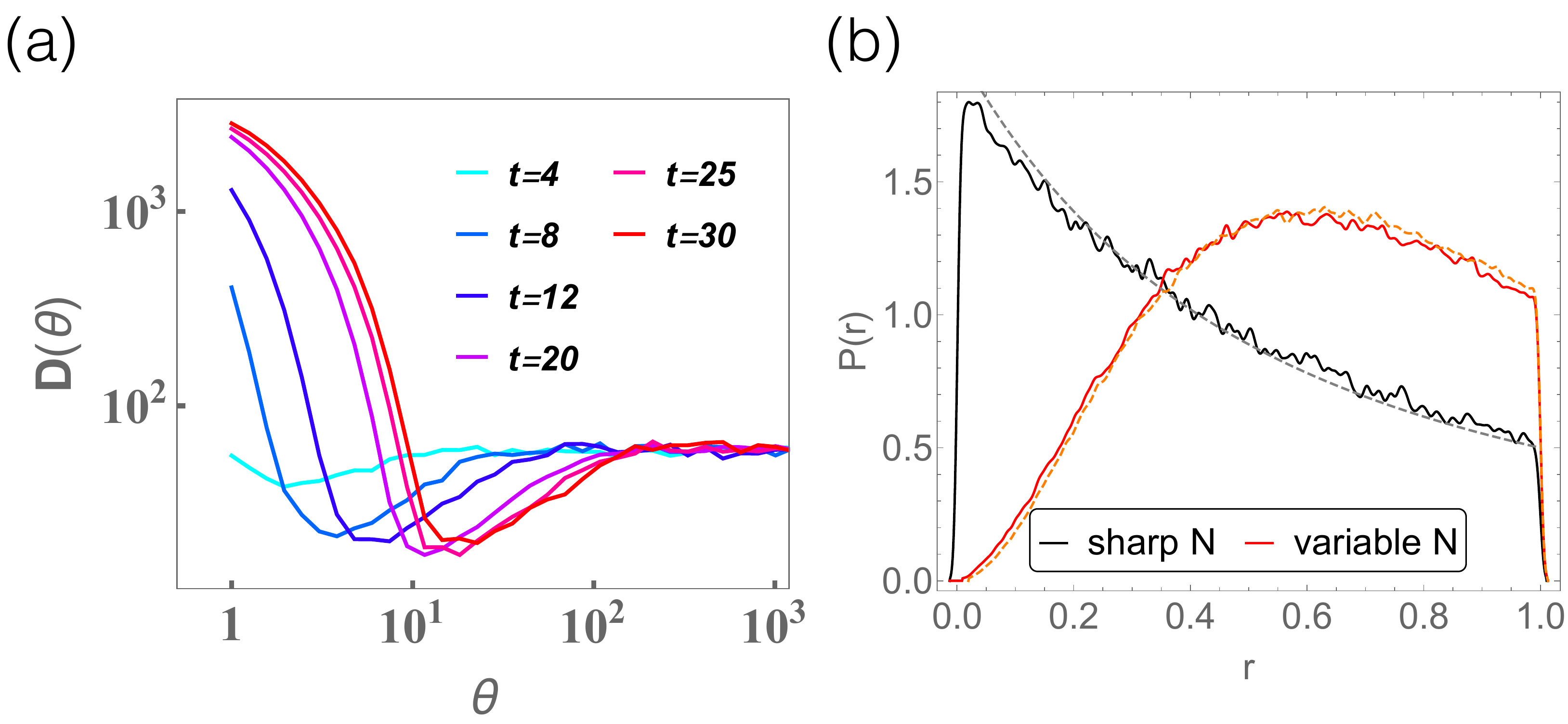}
\caption{ {\it Random unitary circuits with a single conservation law.} (a)~Evolution of the ESFF in the conserving case for an initial state with definite particle number, at $L = 16, l_A = 6$, averaged over 600 samples. Note the appearance of ramp-plateau structure despite the Poisson level statistics in (b). (b)~Level statistics parameter $r$ for the conserving circuit with fixed- and variable-number initial states, which respectively approach Poisson and random-matrix behavior (dashed lines show the exact distributions~\cite{Atas2013} of the $r$ ratio for Poisson and GUE distributions respectively).}
\label{cons}
\end{center}
\end{figure}

(B) \textit{Random circuits with a conservation law}.---To test how robust our results are in the presence of structure in the dynamics we turn to the case with a conserved quantity, which we take to be the $z$-component of the spin. For spin-$1/2$ degrees of freedom the most general conserving two-spin gate acts as a random phase on the states $|\uparrow\uparrow\rangle$ and $|\downarrow\downarrow \rangle$, and as a random $2\times 2$ matrix on the space spanned by $|\uparrow\downarrow\rangle, |\downarrow\uparrow\rangle$.
The conserved quantity is $N \equiv \sum_i \sigma^z_i$, i.e., the number of $\uparrow$ spins. 
We consider two separate classes of initial product states: (i) random eigenstates of $N$ (i.e., random binary strings in a fixed $N$ sector) and (ii) random product states, which are superpositions of different $N$ sectors. 
The results are shown in Fig.~\ref{cons}.

For (i) states that are initially random binary strings, many features are different from the random case. First, the Schmidt decomposition is block-diagonal. Each partition of $N$ into $N_A$ ``particles'' in the sub-interval has $N - N_A$ particles in the complement, so $\rho_A$ has no coherence between states of different $N_A$. Different-$N_A$ blocks do not repel each other, so the global level statistics is Poisson [Fig.~\ref{cons}(b)]. Nevertheless, level repulsion persists within each \textit{individual} block, and manifests itself in the ramp-plateau structure of the ESFF [Fig.~\ref{cons}(a)]. The ESFF is sensitive to level repulsion effects beyond nearest-neighbor levels, and is therefore able to detect intra-block structure, unlike the adjacent gap ratio $r$. 

For random product states (ii), by contrast, the behavior is qualitatively very similar to that of random circuits, although there are quantitative differences in entanglement growth rates~\cite{suppmat}. Again, GUE level statistics emerges on a fixed size-independent timescale when the bond dimension of $\rho_A$ is still growing~\cite{suppmat}. The entanglement DOS behaves qualitatively as in the Haar random unitary circuit model 
although its bandwidth grows even wider for conserving dynamics. One might naively have expected level repulsion in the entanglement spectrum to signal chaos in the underlying dynamics; from this perspective, the irrelevance of the conservation law is unexpected. We observe this feature persists even in dynamics that 
is not chaotic at all but localized~\cite{suppmat}. To summarize, for random product states, the presence of a conservation law has no qualitative effect on the evolution of the entanglement spectrum. Only when the initial states are also eigenstates of the conserved charge does one see qualitatively different evolution in the entanglement spectrum.

%Ising model

(C) \textit{Ising model with transverse and longitudinal fields.}--- To test the generality of our results we now turn to Hamiltonian dynamics. We consider the nonintegrable Ising Hamiltonian [Eq.~\eqref{eqn:HI}] and time-evolve starting from a random product state.
We consider the total system $L=16$ with the subsystem size $l_A=8$. 
We observe the same scale-free $1/f$ probability distribution 
of the eigenvalues of the reduced density matrix \cite{suppmat} [Fig.~\ref{fig1} (a)] and find that
the adjacent gap ratio [Fig. \ref{fig1} (b)] approaches the GUE value on a $o(1)$ time scale. In addition we find the entanglement bandwidth grows for times $t< l_A/2$ and then shrinks at late times. Distinct from RUCs, the entanglement bandwidth starts from a non-zero initial value because the RDM is full rank for Hamiltonian dynamics (since the light-cone set by the Lieb-Robinson bounds is not strict but has exponential tails). 
Lastly, the entanglement bandwidth shrinks well before the entropy saturates [Fig. \ref{fig1}(d)]. In summary, we have obtained all the same features we have observed in purely random circuits.

\begin{figure}[htbp]
\begin{center}
\includegraphics[width=0.45\textwidth]{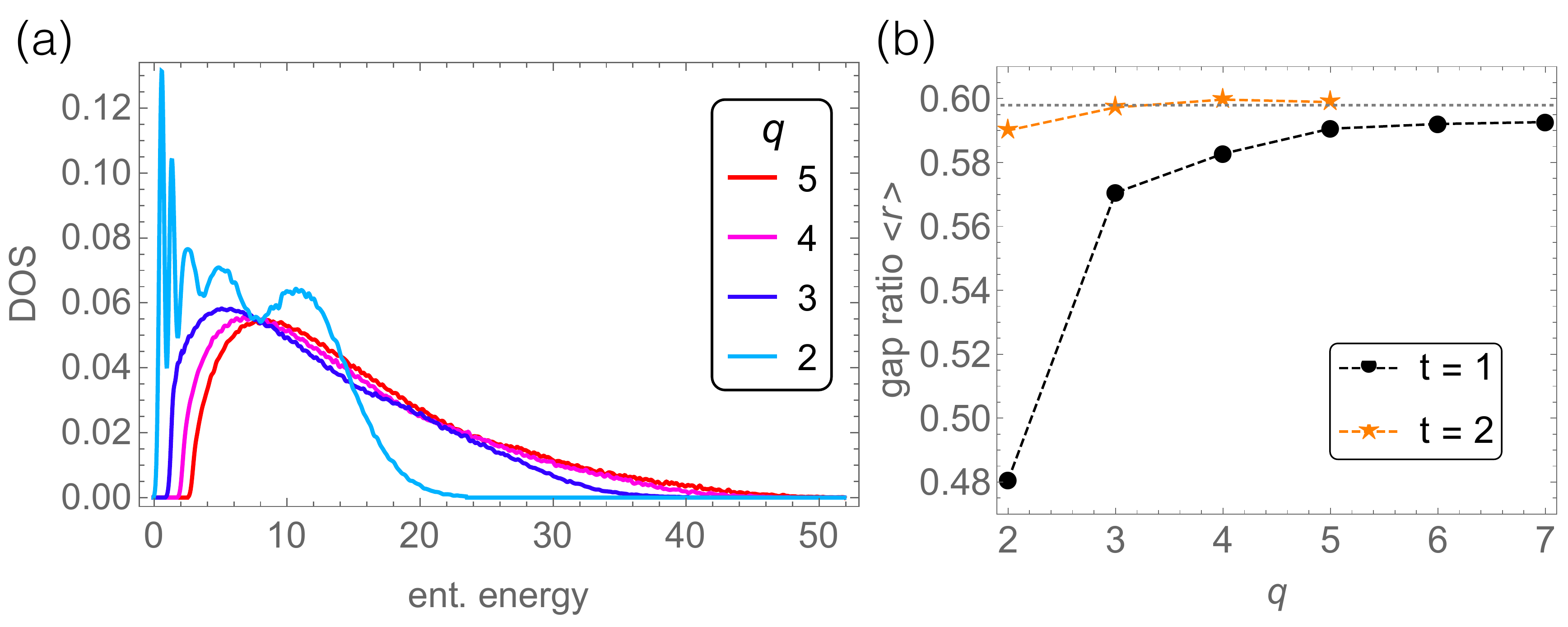}
\caption{{\it Dependence on the local Hilbert space $q$.} (a) Entanglement DOS at a fixed time, $t = 2$, as a function of local Hilbert space dimension $q$. The shape of the DOS does not seem to change much with $q$, though the average entanglement ``energy'' goes down as one might expect. (b) Adjacent gap ratio $r$  vs. $q$ at a fixed time $t = 1$ and $t=2$, these are the only times at which there are appreciable deviations from GUE level statistics for $q>2$ (the dashed line marks the exact GUE value $r\approx 0.599$).}
\label{qdep}
\end{center}
\end{figure}

\textit{Dependence on local Hilbert space}.---
Besides the comparison between different RUCs and the spin Hamiltonian, we compare our results for different dimensions of local Hilbert space $q> 2$,
focusing on purely random circuits (Fig.~\ref{qdep}.) Surprisingly, the entanglement DOS stays broad for all the $q$ we have considered [Fig.~\ref{qdep} (a)]; despite 
the expectation that 
this quantity 
narrows as $q\rightarrow \infty$,
we see no clear sign of narrowing. 
Thus, the approach to the known $q\rightarrow \infty$ behavior is slow; our results suggest that the limit might also be singular.
Turning to the gap ratio $r$ we find [Fig.~\ref{qdep} (b)] that for $q \geq 6$ one has GUE statistics in the entanglement spectrum for $t = 1$. Thus, at large $q$, the onset of level repulsion in the entanglement spectrum is essentially instantaneous.

\textit{Discussion}.---Our results can be qualitatively understood~\footnote{F. Pollmann, private communication.} by invoking the relation between entanglement and operator spreading~\cite{ho2017, Keyserlingk2017, lamacraft2018, knap2018}, as follows: one can expand the reduced density matrix in a basis of strings of Pauli matrices, and study the evolution of these strings in the Heisenberg picture. Strings initially localized on either side of the cut spread out, under time evolution, to more complicated operators that straddle the cut. 
Under the partial trace, most such operators vanish. While the unitary evolution of strings is rank-preserving, the partial trace ``dephases'' components of the reduced density matrix and thereby increases its rank. Heuristically, operators with a given amplitude, when traced out, generate entries of that amplitude in the reduced density matrix. At early times the density matrix is low-rank, so adding a new entry of some size almost always creates a new eigenvalue of the same size.
This picture qualitatively captures the entanglement DOS and level statistics. In RUCs, the speed of the strict causal light-cone ($v_{LC} = 2$) exceeds the butterfly velocity $v_B$ at which generic operators spread. Thus, under time evolution, terms that extend beyond the operator front but within the causal light-cone get generated with small amplitude; those closest to the light-cone are generated at time $t$ with amplitude $\exp\{-[t (v_{LC} - v_B)]^2/(Dt)\}$~\cite{Keyserlingk2017, Nahum2017, xu2018a}, where $D$ is the rate at which the front broadens. 
These exponentially small-amplitude operators generate correspondingly small eigenvalues in the reduced density matrix, leading to entanglement energies that grow linearly in $t$ and thus accounting for the observed {linear} bandwidth expansion. Once the light-cone hits the edge of the subsystem, the density matrix is full rank, and tracing out further operators cannot create new eigenvalues, but instead redistributes weight among existing eigenvalues, causing the spectrum to narrow. The entanglement level statistics can be understood in similar terms: operators that contribute nonzero Schmidt coefficients are those that have crossed the entanglement cut; by virtue of this property they all have overlapping support and are in causal contact. Therefore it is natural for the corresponding eigenvalues to have the statistics described by the random matrix theory~\cite{rgpp}.

Although we presented this argument for RUCs, it can straightforwardly be adapted to Hamiltonian dynamics. The density of states and level statistics of the entanglement spectrum behave qualitatively the same as with RUCs. The main difference is that the reduced density matrix is always full-rank so $S_0$ is not physically relevant. However, if one ``regularizes'' $S_0$ to include only eigenvalues above a certain threshold (that is well above numerical precision), the resulting evolution is qualitatively the same as in RUCs.

\begin{acknowledgments}

 P.-Y. C. thanks M. Tezuka and C.-T. Ma for discussions. S.G. thanks A. Lamacraft, S. Parameswaran, F. Pollmann, and T. Rakovszky for discussions and collaborations on related topics.
P.-Y.C. was supported by the Rutgers Center for Materials Theory postdoctoral grant and 
Young Scholar Fellowship Program by Ministry of Science and Technology (MOST) in Taiwan, under MOST Grant for the Einstein Program MOST
108-2636-M-007-004. X.C. was supported by postdoctoral fellowships from the
Gordon and Betty Moore Foundation, under the EPiQS initiative, Grant GBMF4304, at
the Kavli Institute for Theoretical Physics.
S.G. acknowledges support from NSF Grant No. DMR-1653271. S.G. and J.H.P. performed part of this work at the Aspen Center for Physics, which is supported by NSF Grant No. PHY-1607611, and at the Kavli Institute for Theoretical Physics, which is supported by NSF Grant No. PHY-1748958.  The authors acknowledge the Beowulf cluster at the Department of Physics and Astronomy of Rutgers University and the Office of Advanced Research Computing (OARC) at Rutgers, The State University of New Jersey (http://oarc.rutgers.edu) for providing access to the Amarel cluster and associated research computing resources that have contributed to the results reported here.

\end{acknowledgments}

\bibliographystyle{apsrev4-1}
\bibliography{references}

\clearpage

\pagebreak

\newpage

%%%%%%%%%% Prefix a "S" to all equations, figures, tables and reset the counter %%%%%%%%%%
\setcounter{equation}{0}
\setcounter{figure}{0}
\setcounter{table}{0}
\setcounter{page}{1}
\makeatletter
\renewcommand{\theequation}{S\arabic{equation}}
\renewcommand{\thefigure}{S\arabic{figure}}
\renewcommand{\bibnumfmt}[1]{[S#1]}
\renewcommand{\citenumfont}[1]{S#1}
%%%%%%%%%% Prefix a "S" to all equations, figures, tables and reset the counter %%%%%%%%%%

\begin{widetext}
\begin{center}
\noindent{\Large Supplementary Material for ``Evolution of entanglement spectra under random unitary dynamics". }
\end{center}

In this document we present additional data on the geometry of the random circuits we consider, 
the evolution of the entanglement spectrum in localized systems, 
the evolution of level statistics with time in the various models, 
entanglement dynamics in random conserving circuits and the nonintegrable Ising model, 
the entanglement-energy dependence of the gap ratio, and the reduced density matrix spectral form factors.

\section{Random circuit models}

The random unitary circuits (RUCs) are constructed from the random unitary matrices as shown in Fig. \ref{RCM_fig}.
The form of the time evolution operator can be written as $U(t) = U(t,t-1)U(t-1,t-2) \cdots U(1,0)$, where
\begin{align}
U(t',t'\!-\!1)\!=\!&\bigotimes_{i \in 2 \mathbb{Z}}U_{i,i+1}\! (t',t'\!-\!1) \!\!\!\bigotimes_{i \in 2 \mathbb{Z}+1}\!\!\!\!U_{i,i+1}\! (t',t'\!-\!1),
\end{align}
with $i$ being the site index and $U_{i,i+1}$ are unitary matrices.
We discuss two types of RUCs in the main text. The first one is choosing the 
 unitary matrices with Haar measure. The second one is choosing the conserving random unitaries as in Ref.~\cite{Khemani2017}, which conserve the number of up spins in the $Z$ basis.
 In this supplementary material, we will discuss %the 
 %other two types. One has the similar structure as Fig.\ref{RCM_fig}.
%The unitary matrices are chosen to be Free-fermion random unitaries, as in Ref.~\cite{gullans}, consisting purely of partial-swap gates. %\jp{NEED TO ADD CHALKER MODEL}
%In this case, it yields a free-fermion Floquet model, which is Anderson localed~\cite{Agarwal-2017}; the other cases thermalize. Unless otherwise specified we take $q = 2$.
%The other circuit model we considered is 
a Floquet model with a different circuit geometry~\cite{cdc_prl}, which is believed to be many-body localized~\cite{cdc_prl}; our main results hold for this model as well.

\begin{figure}[!htbp]
\begin{center}
\includegraphics[width=0.35\textwidth]{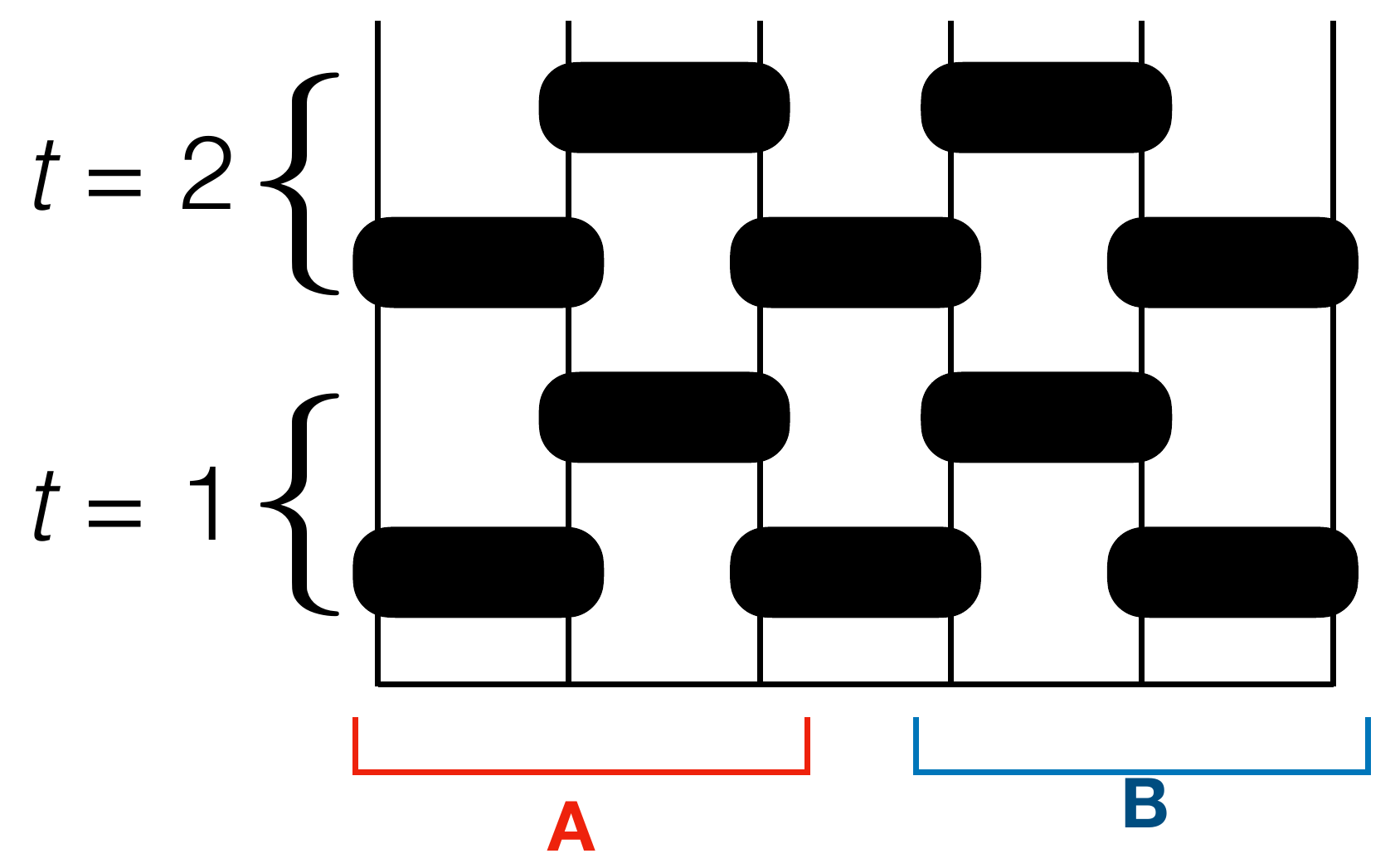}
\caption{Depiction of the random unitary circuits considered here: one time-step involves applying random gates on the even bonds, then on the odd bonds.}
\label{RCM_fig}
\end{center}
\end{figure}

\subsection{Many-body localized model}

%The free-fermion model exhibits some distinctive features, and it is natural to ask if these are also present in other localized models. To this end 
We first consider the localized model introduced in Ref.~\cite{cdc_prl}, which was argued to be in the many-body localized phase for $q = 2$. This model consists of two types of gates: a gate that purely adds phases $\eta \in [-\phi, \phi]$ in the $Z$ basis, and a gate consisting of random single-site rotations. Thus its one-cycle time evolution operator has the form

\begin{equation}
U = \prod_{i} \exp(i \eta \sigma^z_i \sigma^z_{i+1}) \prod_i R_i,
\end{equation}
where $R_i$ is a random single-site gate. 
%\jp{JP: SG CAN WE WRITE THE MODEL OUT WE HAVE SPACE HERE?}. 
These gates are applied periodically in time giving rise to a Floquet system; the model has the nice property of having a controllable parameter $\phi$, which tunes between decoupled and strongly coupled qubits. Note that because of the different circuit geometry the light-cone is slower in this model than in the others we have considered: specifically, $v_{LC} = t$. 

We find that the dynamics of the entanglement entropy for this model is consistent with many-body localization demonstrating that $S_2 \sim \log t$, which has a clear log growth with time (Fig.~\ref{chalker}). However, the ratio $r$ behaves as it does in the other models discussed in the main text. Accounting for the fact that all timescales are doubled, $r$ appears to saturate to GUE on the timescale we would have predicted from the other models ($t \simeq 8$) -- see Fig.~\ref{chalker}, middle. The entanglement bandwidth also grows rapidly with time, again consistent with the behavior seen in chaotic models (Fig.~\ref{chalker}, right). This might seem counterintuitive but is actually what our analysis in terms of operator spreading would predict: the light-cone speed is now $1$ while the butterfly speed is zero (since we are in the localized phase), but as we argued in the main text the bandwidth growth is set by $|v_B - v_{LC}|$ so it is finite in this case. Finally, the existence of level repulsion follows once again from the fact that all the operators that contribute to entanglement are in causal contact with one another, since they have all traversed the entanglement cut. 

The entanglement entropy itself (as well as R\'enyi entropies with $\alpha > 1$) are dominated by the top few Schmidt coefficients, which indeed behave dramatically differently in chaotic and localized systems. However, the measures we are looking at here concern the \emph{bulk} of the entanglement spectrum, and this appears qualitatively insensitive to whether the dynamics is actually chaotic.
%We have not seen any sign of non-monotonicity in the evolution of $r$ for smaller cuts, however. 

\begin{figure}[htbp]
\begin{center}
\includegraphics[width=0.9\textwidth]{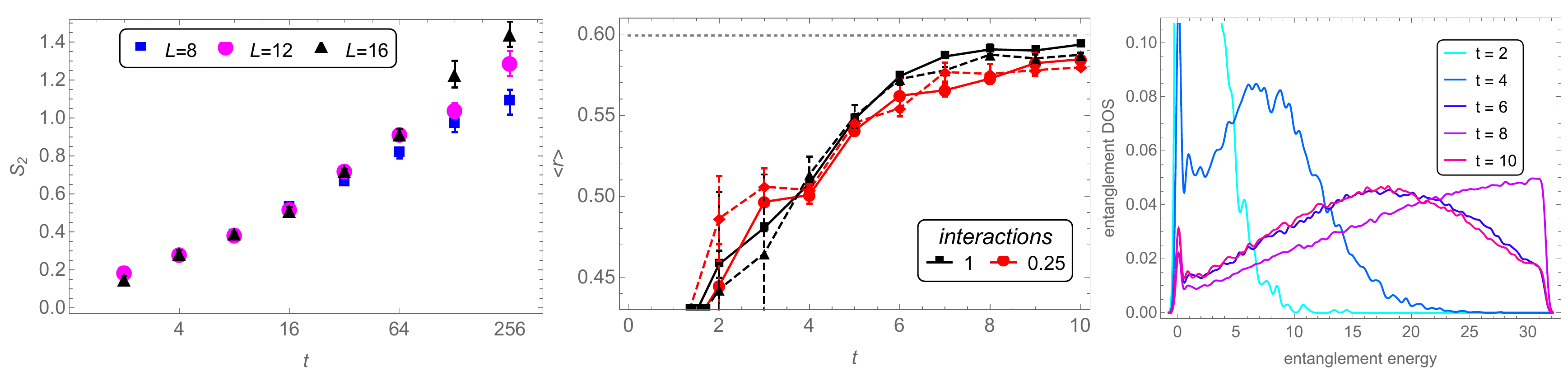}
\caption{Entanglement dynamics of the localized model of Ref.~\cite{cdc_prl}. Left: Renyi entropy $S_2$ vs. time for $\phi = 1$; results are roughly consistent with logarithmic growth. Center: evolution of $r$ for a half-system cut, at various values of $\phi$ (labelled as interactions in the legend); solid lines are for $L = 16$ and dashed lines for $L = 12$. Note the non-monotonic behavior at $L = 12$ which is absent at the larger system size. Right: time-evolution of entanglement DOS for this model at $L = 16$, averaged over $500$ samples.}
\label{chalker}
\end{center}
\end{figure}

%Our results for this model are broadly consistent with those for the free-fermion localized model. Indeed the phenomena appear rather more strongly: even half-system cuts exhibit the non-monotonic behavior of the adjacent gap ratio in this model. 

\subsection{Summary of $\langle r \rangle(t) $ for various models}

In Fig.~\ref{giantfig} we summarize the data we have collected for the evolution of the nearest-neighbor adjacent gap ratio for a variety of models. Evidently, $\langle r \rangle$ approaches its saturation value on very similar timescales in all of these models, despite their considerable differences. In addition to the models for which we have shown data, we have observed similar behavior for Floquet circuits in which the same random gates are applied at every timestep, as well as for random free fermion models. 

\begin{figure}[htbp]
\begin{center}
\includegraphics[width = 0.6\textwidth]{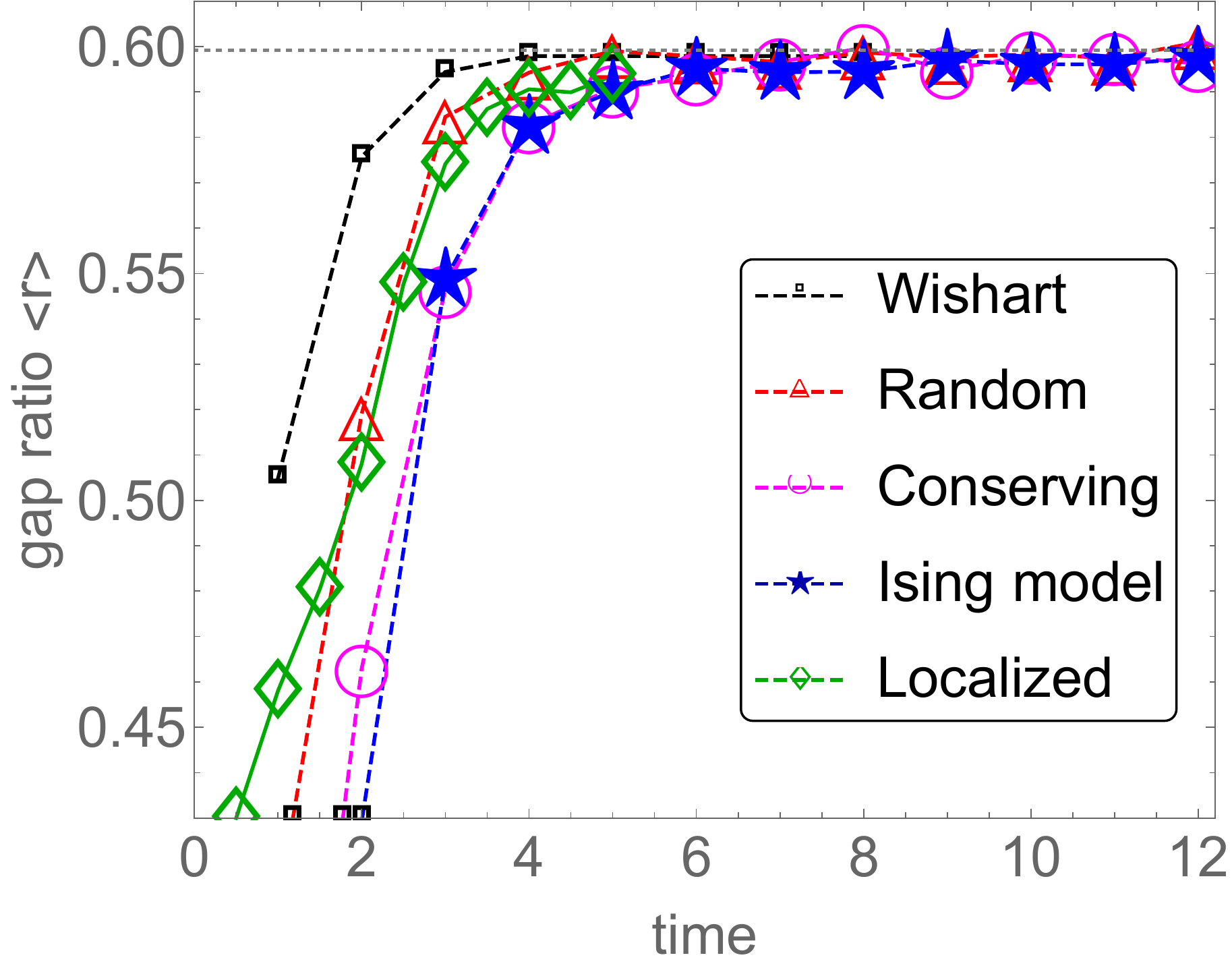}
\caption{Summary of the evolution of $\langle r \rangle (t)$ for a variety of models, compared with Wishart random matrices of size set by the bond dimension of the corresponding circuit at time $t$. Note the essentially model-independent saturation timescale, which occurs in both chaotic and localized systems.}
\label{giantfig}
\end{center}
\end{figure}

\section{Further details on conserving circuits}

\begin{figure}[htbp]
\begin{center}
\includegraphics[width = 0.35\textwidth]{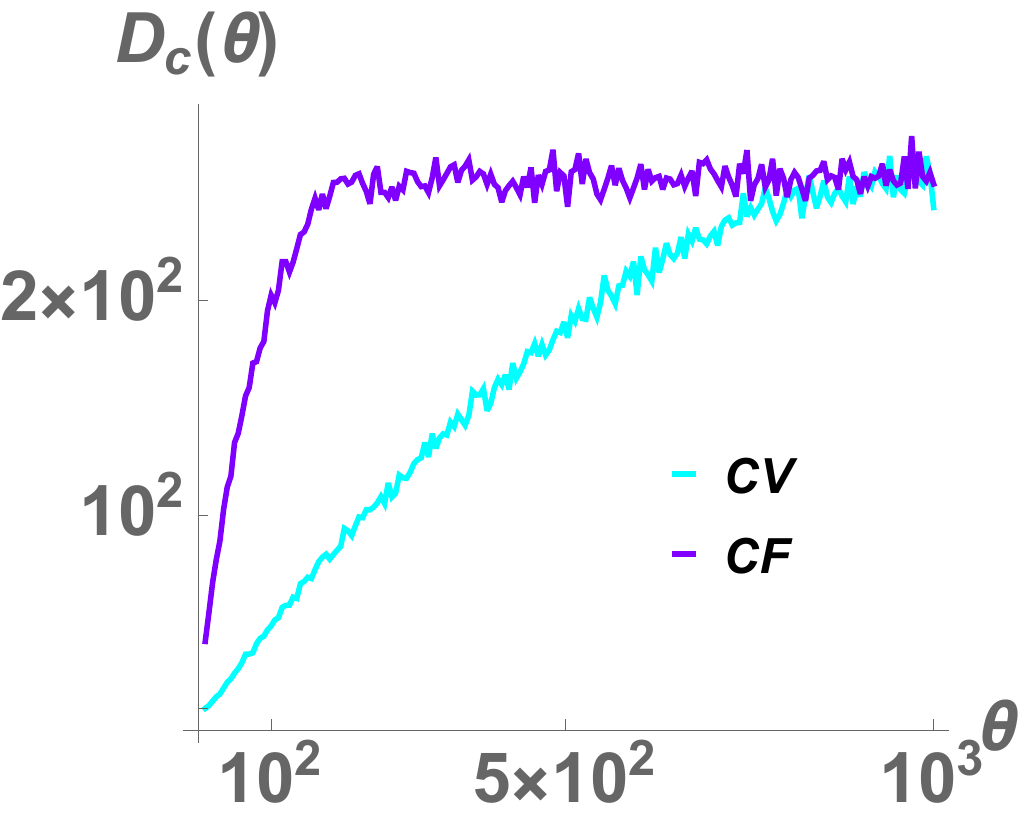}
\includegraphics[width = 0.4 \textwidth]{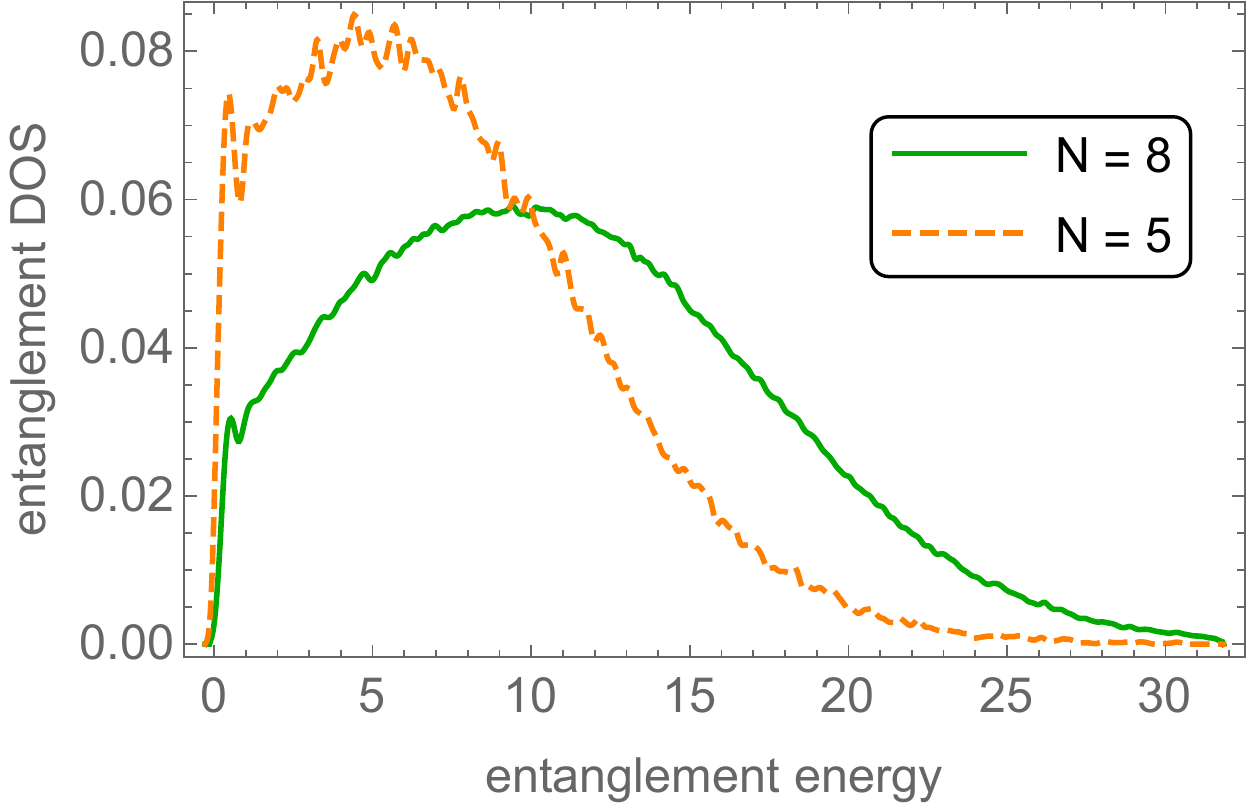}
\caption{Left: connected ESFF $D_c(\theta)$ for $t = 20$ on a \textit{linear} scale for the conserving circuit with fixed and variable-number initial states (CF and CV respectively); both exhibit ramps but the latter is more obviously linear. Right: entanglement DOS in different number sectors for a fixed-number initial state.}
\label{fig85}
\end{center}
\end{figure}

While the ESFF clearly shows ramps in the conserving circuits, it is not clear on the logarithmic scale that these ramps are truly linear. To address this we have plotted the connected ESFF on a linear scale (Fig.~\ref{fig85}). In the case of a random initial product state the ramp is manifestly linear. For a fixed particle number (the case that gave Poisson level statistics) the linearity of the ramp is less manifest; however, the observed behavior is consistent with a linear ramp at early times with a slope that decreases at late times. 

One might also wonder whether our results for the entanglement DOS in the conserving circuit, fixed-number case are qualitatively modified by averaging over different particle-number sectors. To address this we break out results by particle-number sector in the right panel of Fig.~\ref{fig85}. We find that the EDOS is qualitatively the same in different number sectors, though the entanglement bandwidth is largest for half-filling.

%Lastly, we compare the growth of the entanglement entropy for the purely random circuit and the conserving circuit in Fig. \ref{cons_2_2}(a).
%The slope of the entanglement entropy in the random circuits with $S_z$ conservation law is smaller than the random circuits without conservation laws as expected.
%Although many similarities shared in purely random circuit and the random circuits with $S_z$ conservation law.
%The difference between these two cases is also manifest in the qualitative shape of the entanglement DOS [Fig.~\ref{cons_2_2}(b)], which is manifestly different from that in random circuits. 
%We have checked that the DOS has this shape even within single particle-number sectors.

%\begin{figure}[tb]
%\begin{center}
%\includegraphics[width=.7\textwidth]{figcons2_2}
%\caption{(a)~Entanglement growth for conserving vs. random unitaries, for $L = 20, l_A = 10$, averaged over 100 samples. (b)~Entanglement DOS at a fixed time $t = 4$ for three different cases: random (R), conserving with fixed particle number (CF), and conserving with a random product state (CV). Cases R and CV are similar, but distinct from CF.}
%\label{cons_2_2}
%\end{center}
%\end{figure}

\section{The Ising model with transverse and longitudinal fields}

The Ising model with transverse and longitudinal fields has the following Hamiltonian
\begin{align}
H=\sum_i J \sigma^z_i\sigma^z_{i+1}+h_x\sum_i \sigma^x_i +h_z\sum_i\sigma^z_i
\end{align}
This model is chaotic when $h_z\neq 0$ and in the calculation, we take the parameter $(h_x/J,h_z/J)=(0.9045, 0.809)$~\cite{Kim-2013,Kim-2015} (we set $J=1$ as the unit of energy). For an initial random product state, under Hamiltonian dynamics, the reduced density matrix will eventually thermalize with the effective temperature specified by the energy of the initial state. In our calculation, we take the energy to zero and we find that the ESFF starts to form a ramp when $t>6$ at which the adjacent gap ratio  saturates to the GUE value $\langle r\rangle\approx 0.599$. After sufficient time evolution, the ESFF develops a ramp-plateau structure, the same as we observe in the Page state, signaling level repulsion among the entanglement energy levels (Fig.~\ref{fig085}).

\begin{figure}[htbp]
\begin{center}
\includegraphics[width = 0.3\textwidth]{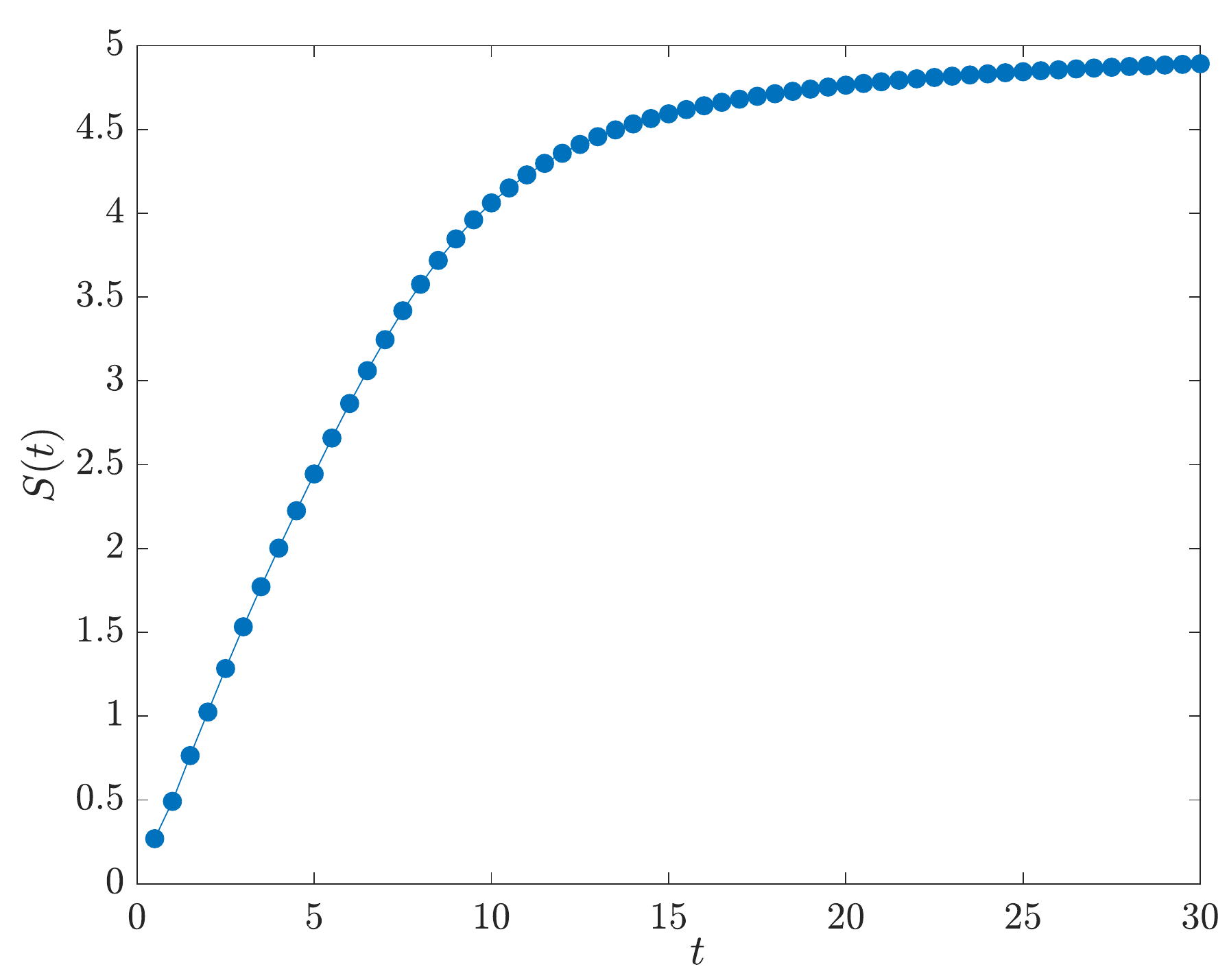}
\includegraphics[width = 0.3 \textwidth]{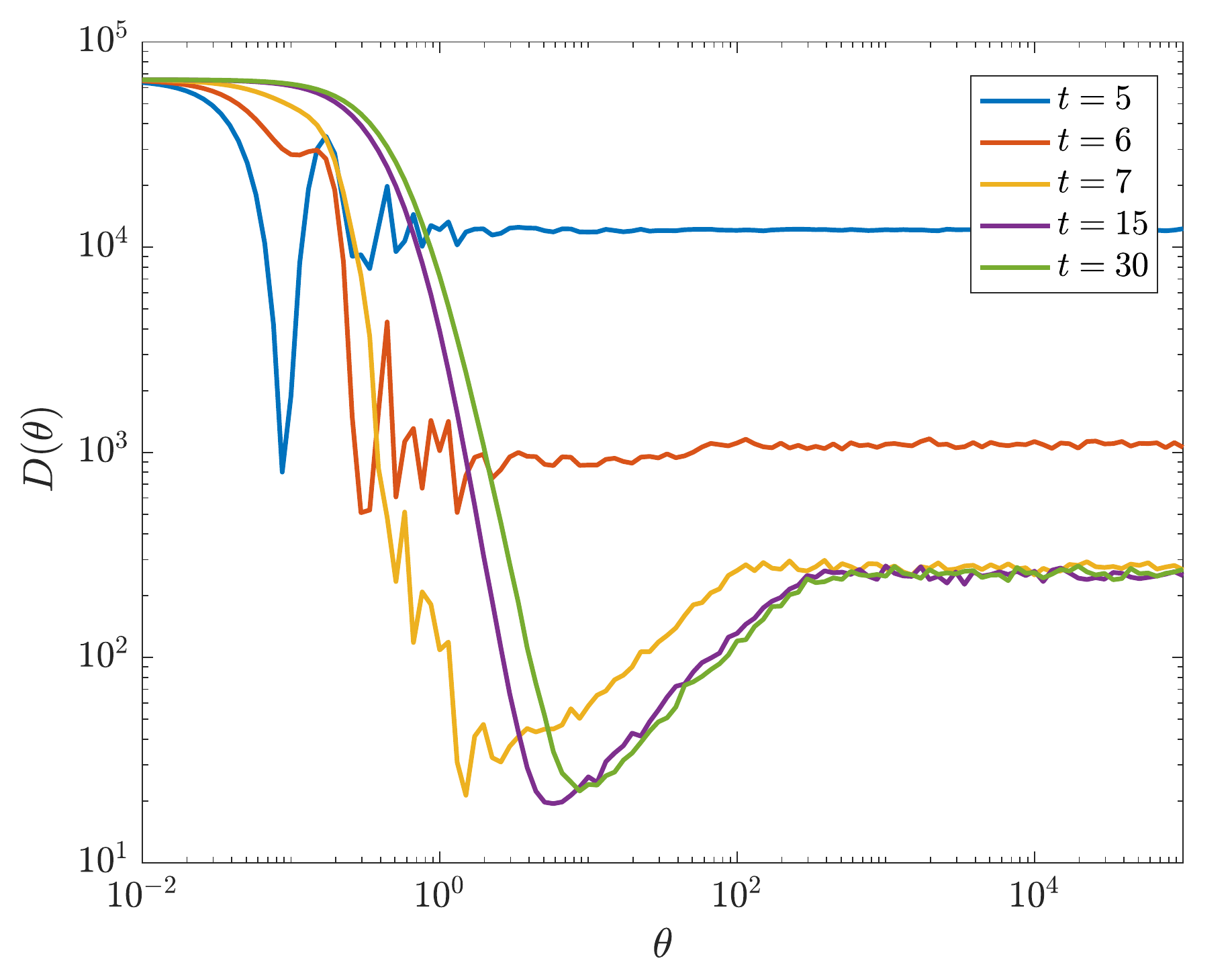}
\includegraphics[width = 0.315 \textwidth]{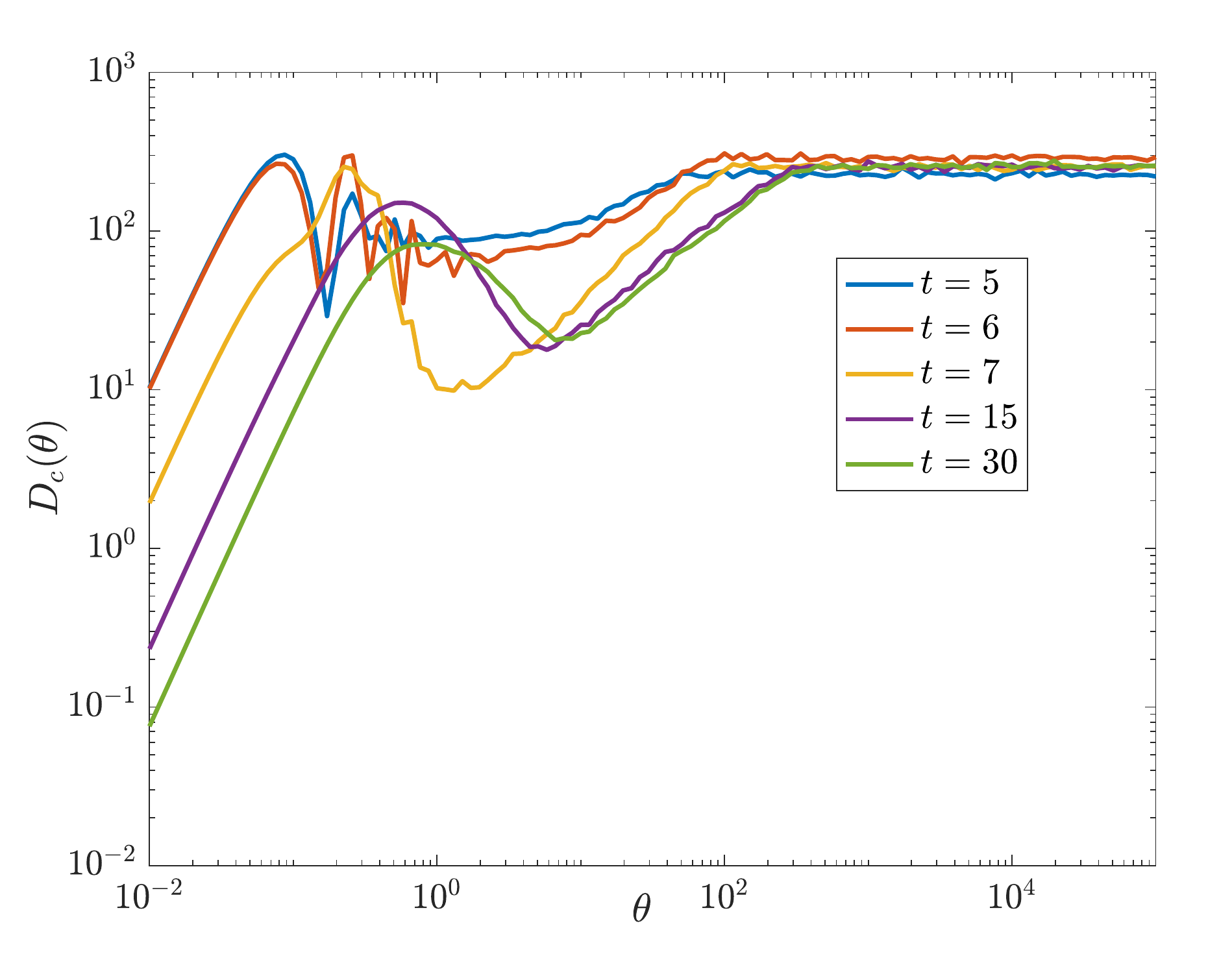}
\caption{(a)~Von Neumann entanglement entropy $S(t)$ as a function of time $t$ for Ising Hamiltonian. The initial state is random product state with $E=0$. The system has $L=16$ with $L_A=8$. (b)~ESFF at different time. The ramp starts to develop when $t>6$. (c)~ The connected ESFF at different time.}
\label{fig085}
\end{center}
\end{figure}

\section{Entanglement-energy dependence of the onset of GUE level statistics}

A natural question about the entanglement level statistics concerns its behavior as a function of the entanglement energy. This dependence is illustrated in Fig.~\ref{edep}. The essential trend is that GUE level statistics develops first in regions of high entanglement DOS (away from the edges of the entanglement spectrum) then spreads to the edges. In the random model the deviations are quite small even at the earliest times; by contrast the cases with a conservation law clearly show less random-matrix-like behavior at the edges of the spectrum.

\begin{figure}[htbp]
\begin{center}
\includegraphics[width = 0.8\textwidth]{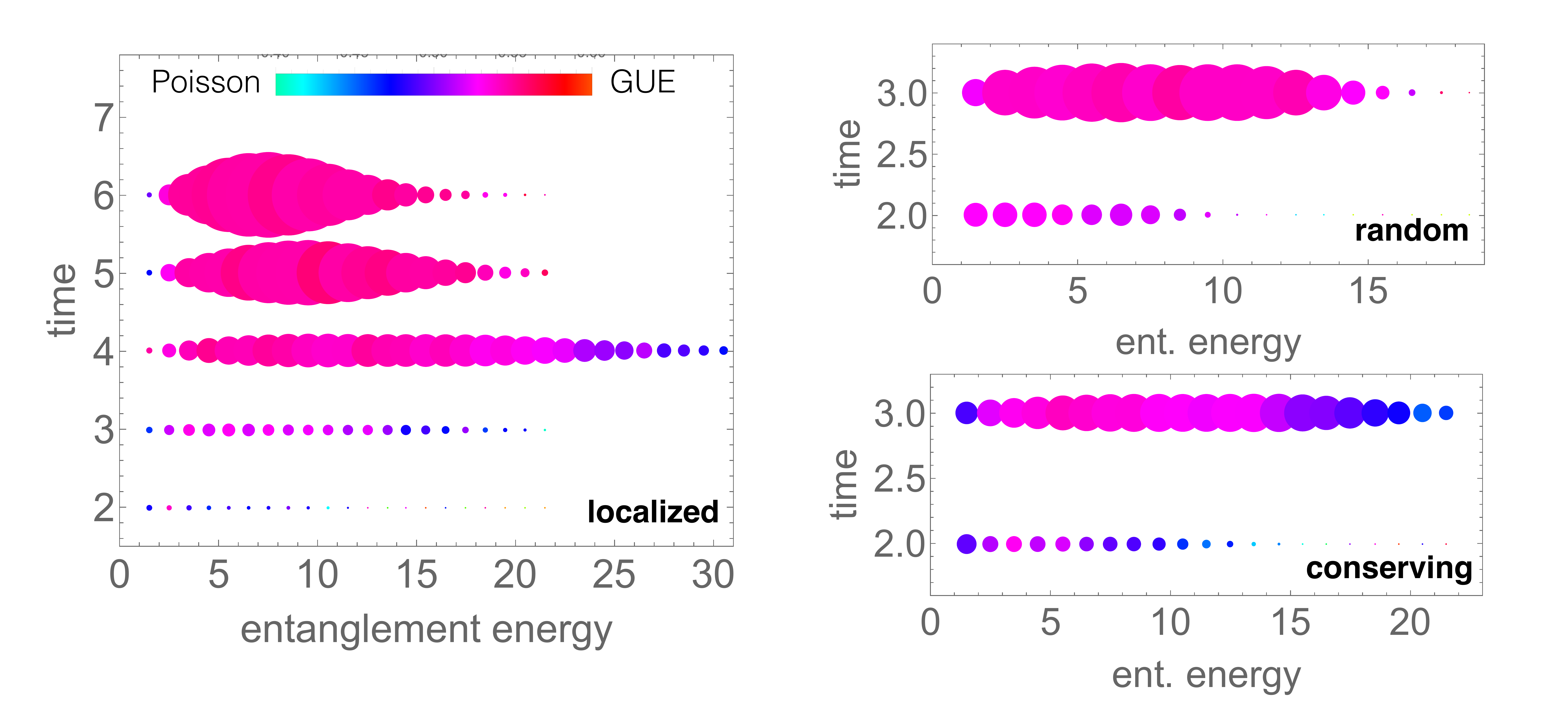}
\caption{Adjacent gap ratio $r$ vs. entanglement energy for various models. The color indicates the value of $r$ averaged over a part of the spectrum, while the size of a dot indicates the entanglement DOS in that bin.}
\label{edep}
\end{center}
\end{figure}

\section{Spectral form factor for the reduced density matrix}
In the main text, we discuss the ESFF which is a characteristic function of the level distributions of the entanglement spectrum.
Here we investigate the spectral form factor for the reduced density matrix, which characterizes the level statistics of the reduced density matrix, expressed as:
\beq
G(\theta) \equiv \left\langle \sum\nolimits_{n,m} e^{i \theta(\lambda_n - \lambda_m)} \right\rangle.
\eeq
where $\lambda_n$ are the eigenvalues of the reduced density matrix.
The connected spectral form factor for the reduced density matrix is defined as
$G_c(\theta) = G(\theta) - \vert\langle \sum\nolimits_{n} \exp(i \theta \lambda_n )\rangle \vert^2$.
We denote these (connected) spectral form factors $\rho$SFF~\cite{Chen2017}.
Fig.~\ref{formfactors_2} shows the behavior of the $\rho$SFF [$G(\theta)$ and $G_c(\theta)$].
Contrasting Figs.~\ref{formfactors_2} with Fig.~\ref{formfactors}, we find
the ESFF  develops a ramp-plateau structure, signaling level repulsion among large Schmidt coefficients; the $\rho$SFF takes longer to develop the analogous structure. 
This is because the large entanglement energies dominate the dephasing at small $\theta$ in the ESFF, but dominate large-$\theta$ behavior in the $\rho$SFF. 

%{\clb add this discussion in the main text: $D(\theta)$ develops the chaotic ramp-plateau structure at much earlier times, because its large-$\theta$ asymptotics is dominated by large Schmidt coefficients (low entanglement ``energies'').}

 %(i.e., high entanglement energies) dephase at small $\theta$ in the ESFF, but dominate large-$\theta$ behavior in the $\rho$SFF. 

%Lastly, we also compute the connected ESFF $D_c(\theta) = D(\theta) - \vert\langle \sum\nolimits_{n} \exp(i \theta E_n )\rangle \vert^2$ (and $G_c(\theta)$ is defined similarly for the $\rho$SFF).
%These form factors have the advantage of capturing gap correlations beyond nearest neighbor, but the disadvantage of being sensitive to the overall entanglement DOS, which as we have seen are strong [Fig.~\ref{fig1} (c)]. 

\begin{figure}[tb]
\begin{center}
\includegraphics[width=.75\linewidth]{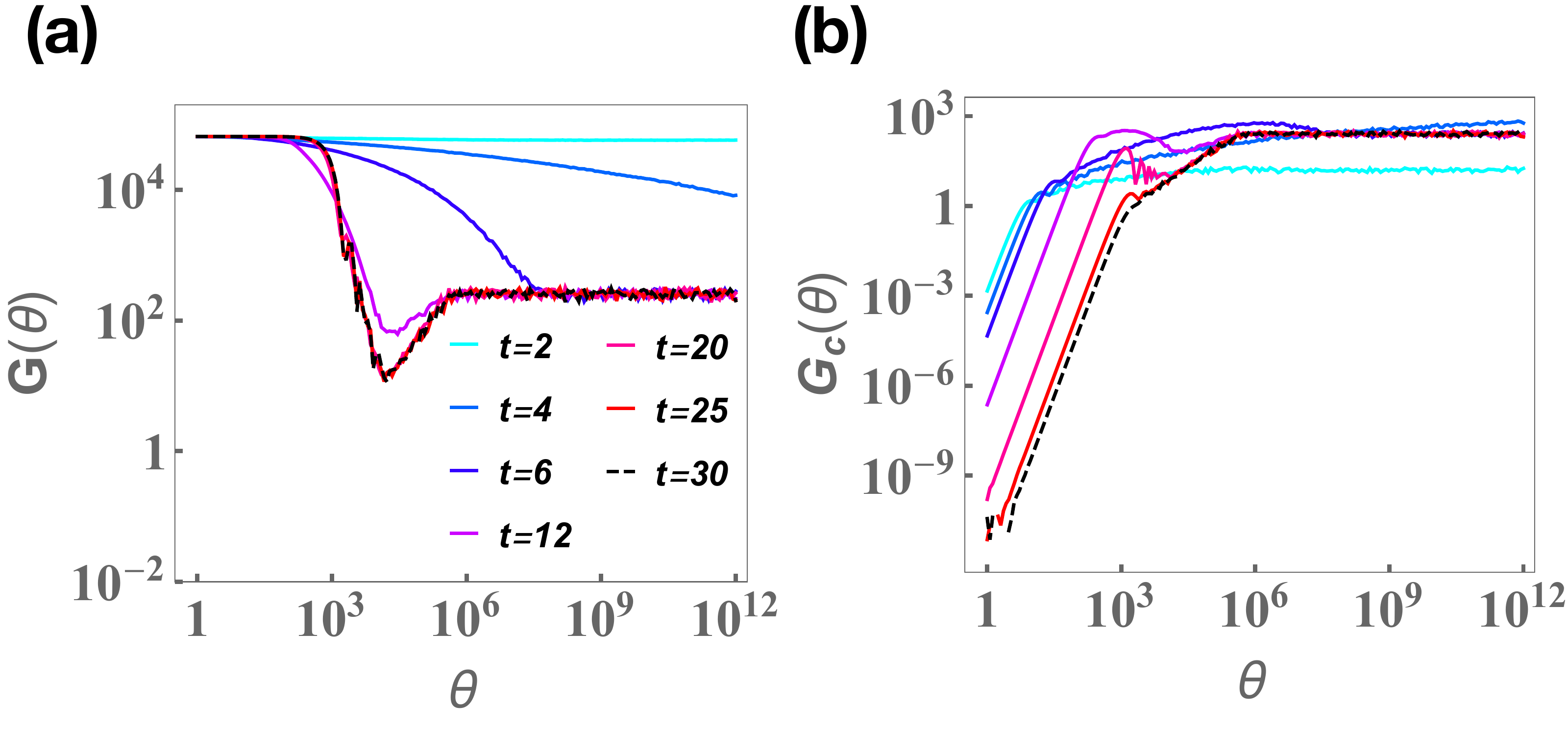}
\caption{The reduced-density matrix (connected) spectral form factor $\rho$SSF
(a) $G(\theta)$ and  (b) $G_c(\theta)$ at various times and the
entanglement (connected) spectral form factor.}
\label{formfactors_2}
\end{center}
\end{figure}
%
%We now present further results on the reduced density matrix spectral form factor $\rho$SFF (see main text for definition) and the connected $\rho$SFF, for the random and conserving models. The connected and full versions are related by
%
%\beq
%G_c(\theta) = G(\theta) - \left\vert \sum_\alpha e^{i q_\alpha \theta} \right\vert^2.
%\eeq
%By subtracting off the initial non-monotonicity, $G_c(\theta)$ reveals more of the ``ramp'' regime in the form factor. 

First, Fig.~\ref{F3} shows the behavior as a function of subsystem size, at a fixed late time $t = 20$. $G_c(\theta)$ has a ramp-plateau structure in all cases, but in $G(\theta)$ the ramp is increasingly hidden at larger subsystems by the initial transient.
\begin{figure}[!htbp]
\center
  \includegraphics[width=0.6 \textwidth] {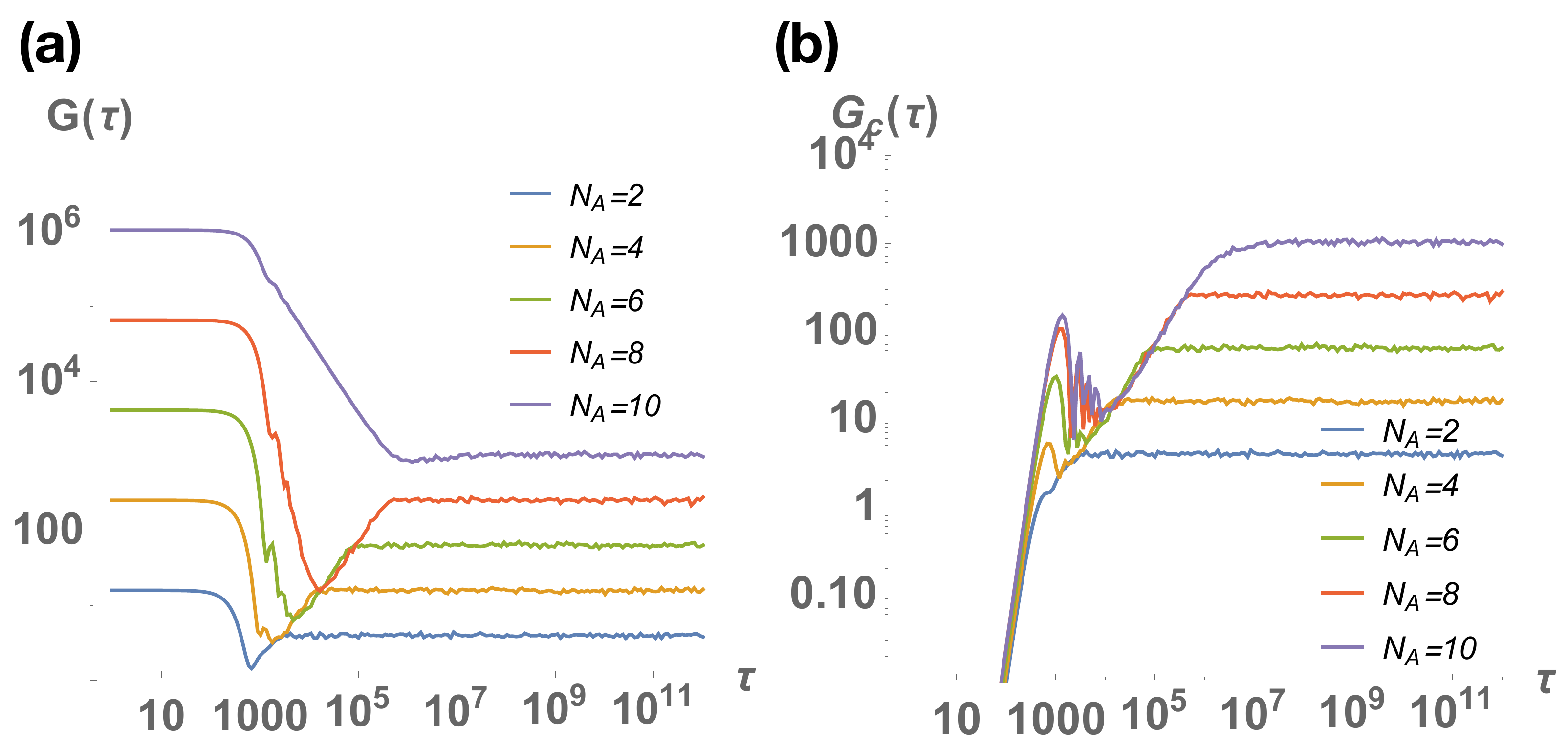}
  \caption{(a) The reduced density matrix spectral form factor and (b) connected spectral form factor at time $t=20$ and total system $N=20$ with $N_A=2,4,6,8,10$.}
  \label{F3}
\end{figure} 
Next, we consider the conserving model. Figs.~\ref{F6} and~\ref{F7} show the behavior of the $\rho$SFF for various subsystem sizes and various times. These data are for a random initial product state, i.e., not a number eigenstate. The behavior is essentially identical to that seen in the purely random circuit.

\begin{figure}[!htbp]
\center
  \includegraphics[width=0.6 \columnwidth] {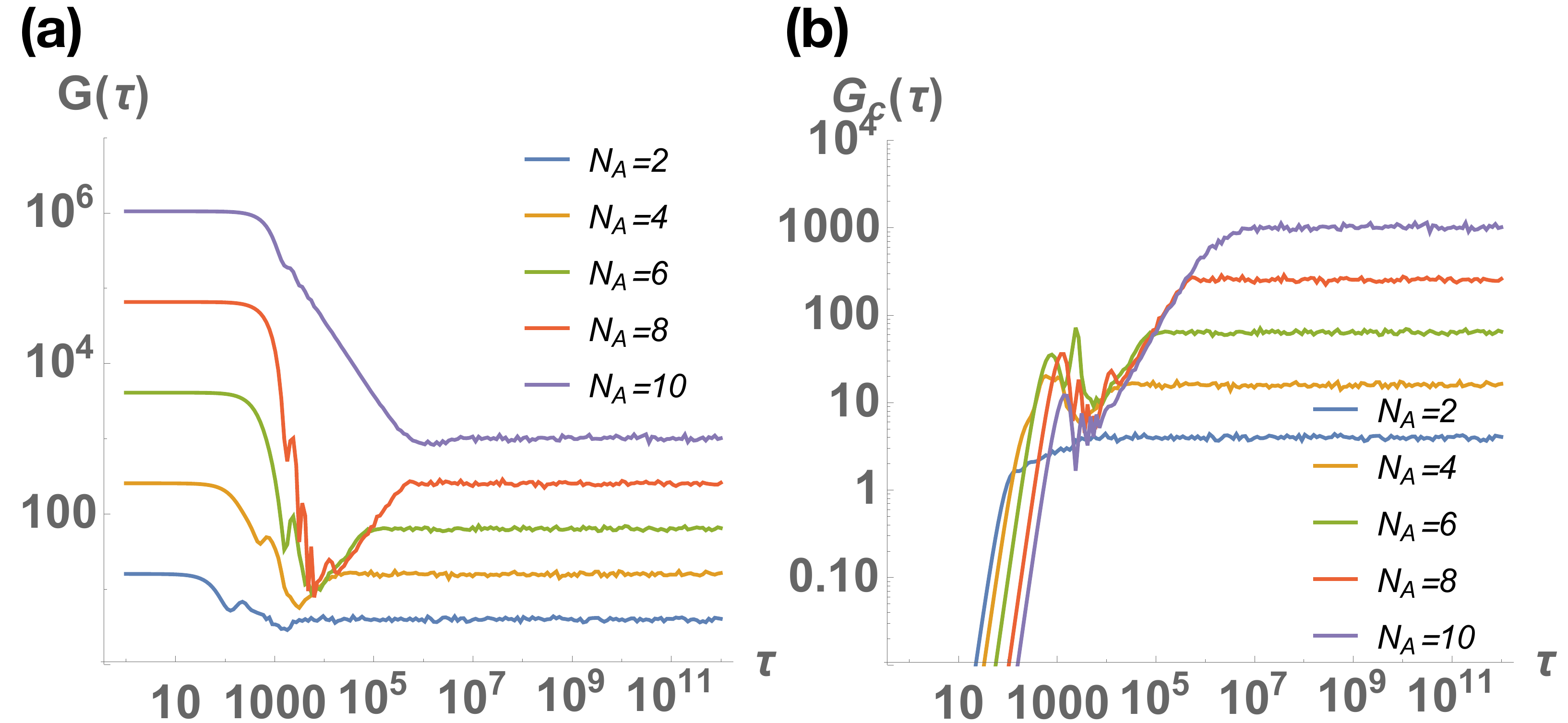}
  \caption{(left) The spectral form factor and (right) the connected spectral form factor for the layer time $l=40$ and total system $N=20$ with $S_z$ conservation and varying the subsystem sizes $N_A=2,4,6,8,10$.}
  \label{F6}
\end{figure}

\begin{figure}[!htbp]
\center
  \includegraphics[width=0.6 \columnwidth] {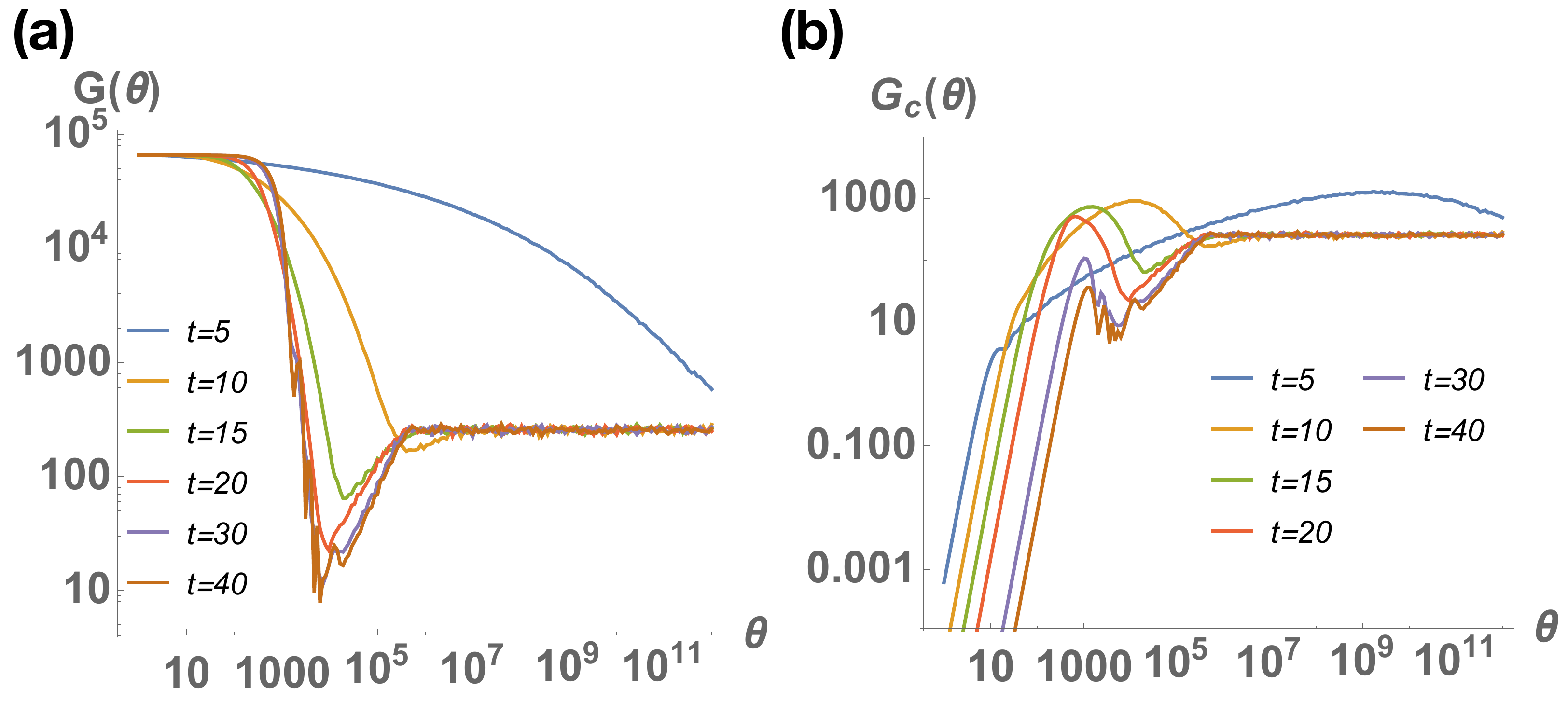}
  \caption{(a) The spectral form factor and (b) the connected spectral form factor for the subsystem $N_A=8$ and total system $N=20$ with $S_z$ conservation and with different times $t=5,10,15,20,30$.}
  \label{F7}
\end{figure} 

Finally, Fig.~\ref{F8} shows the behavior of the $\rho$SFF, as well as the entanglement entropy, for a \emph{specific} state with fixed initial particle number, the Neel state. Even though adjacent level statistics is Poisson in this case, the $\rho$SFF is sensitive to correlations beyond nearest-neighbor levels, and consequently develops a ramp-plateau structure.

\begin{figure}[!htbp]
\center
  \includegraphics[width= 0.6 \columnwidth] {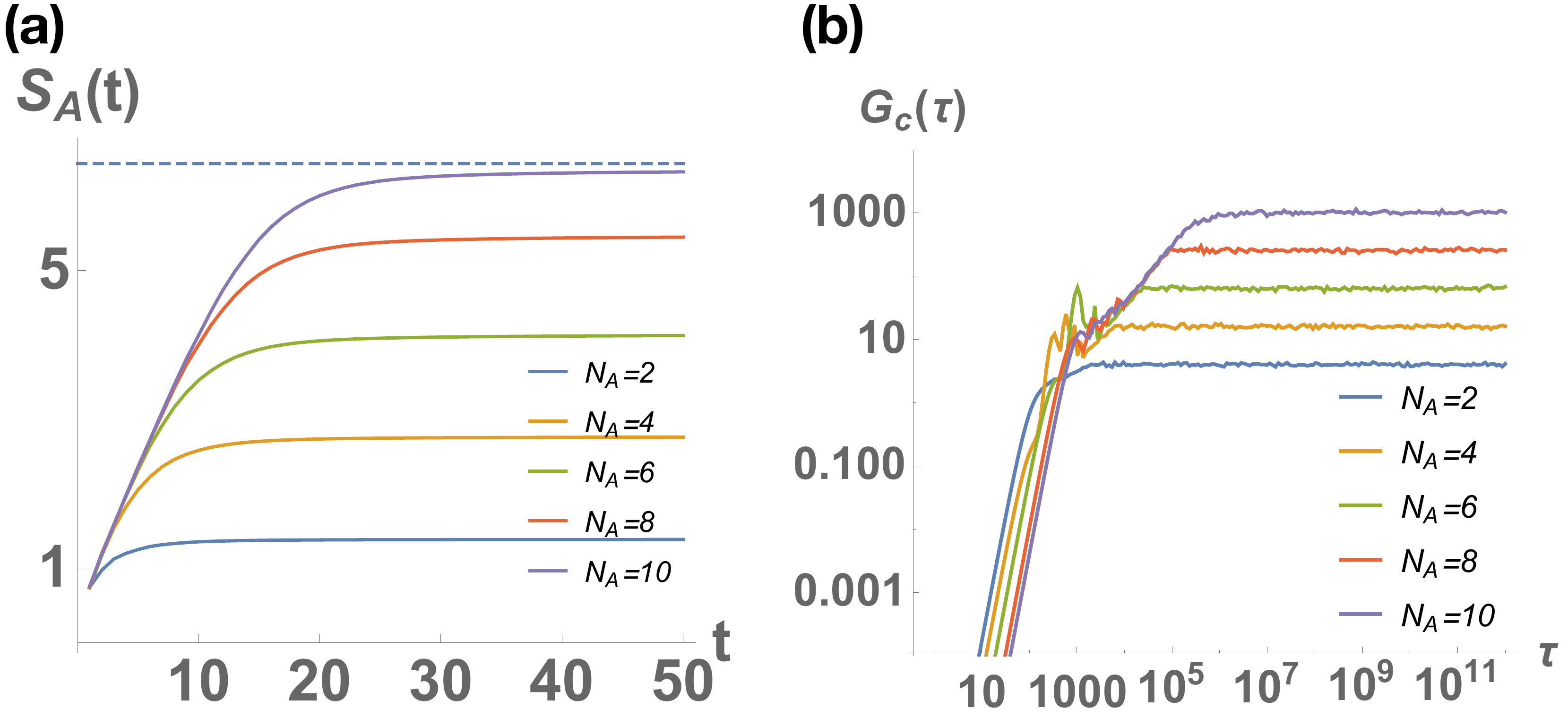}
  \caption{(a) The entanglement entropy as a function of layer time and (b) the connected spectral form factor $G_c(\tau)$  with different subsystem size $N_A$,
   total system size $N=20$ with $S_z$ conservation and with the fixed initial state $|\psi_i\rangle = |\uparrow\downarrow\uparrow\downarrow\cdots \uparrow\downarrow \rangle$.
   The dashed line in (a) is the value of the entanglement entropy for the Page state with $N_A=10$.}
  \label{F8}
\end{figure}

\end{widetext}
\bibliographystyle{apsrev4-1}
\bibliography{references}

\begin{thebibliography}{39}%
\makeatletter
\providecommand \@ifxundefined [1]{%
 \@ifx{#1\undefined}
}%
\providecommand \@ifnum [1]{%
 \ifnum #1\expandafter \@firstoftwo
 \else \expandafter \@secondoftwo
 \fi
}%
\providecommand \@ifx [1]{%
 \ifx #1\expandafter \@firstoftwo
 \else \expandafter \@secondoftwo
 \fi
}%
\providecommand \natexlab [1]{#1}%
\providecommand \enquote  [1]{``#1''}%
\providecommand \bibnamefont  [1]{#1}%
\providecommand \bibfnamefont [1]{#1}%
\providecommand \citenamefont [1]{#1}%
\providecommand \href@noop [0]{\@secondoftwo}%
\providecommand \href [0]{\begingroup \@sanitize@url \@href}%
\providecommand \@href[1]{\@@startlink{#1}\@@href}%
\providecommand \@@href[1]{\endgroup#1\@@endlink}%
\providecommand \@sanitize@url [0]{\catcode `\\12\catcode `\$12\catcode
  `\&12\catcode `\#12\catcode `\^12\catcode `\_12\catcode `\%12\relax}%
\providecommand \@@startlink[1]{}%
\providecommand \@@endlink[0]{}%
\providecommand \url  [0]{\begingroup\@sanitize@url \@url }%
\providecommand \@url [1]{\endgroup\@href {#1}{\urlprefix }}%
\providecommand \urlprefix  [0]{URL }%
\providecommand \Eprint [0]{\href }%
\providecommand \doibase [0]{http://dx.doi.org/}%
\providecommand \selectlanguage [0]{\@gobble}%
\providecommand \bibinfo  [0]{\@secondoftwo}%
\providecommand \bibfield  [0]{\@secondoftwo}%
\providecommand \translation [1]{[#1]}%
\providecommand \BibitemOpen [0]{}%
\providecommand \bibitemStop [0]{}%
\providecommand \bibitemNoStop [0]{.\EOS\space}%
\providecommand \EOS [0]{\spacefactor3000\relax}%
\providecommand \BibitemShut  [1]{\csname bibitem#1\endcsname}%
\let\auto@bib@innerbib\@empty
%</preamble>
\bibitem [{\citenamefont {Deutsch}(1991)}]{deutsch_eth}%
  \BibitemOpen
  \bibfield  {author} {\bibinfo {author} {\bibfnamefont {J.~M.}\ \bibnamefont
  {Deutsch}},\ }\href {\doibase 10.1103/PhysRevA.43.2046} {\bibfield  {journal}
  {\bibinfo  {journal} {Phys. Rev. A}\ }\textbf {\bibinfo {volume} {43}},\
  \bibinfo {pages} {2046} (\bibinfo {year} {1991})}\BibitemShut {NoStop}%
\bibitem [{\citenamefont {Srednicki}(1994)}]{srednicki_eth}%
  \BibitemOpen
  \bibfield  {author} {\bibinfo {author} {\bibfnamefont {M.}~\bibnamefont
  {Srednicki}},\ }\href {\doibase 10.1103/PhysRevE.50.888} {\bibfield
  {journal} {\bibinfo  {journal} {Phys. Rev. E}\ }\textbf {\bibinfo {volume}
  {50}},\ \bibinfo {pages} {888} (\bibinfo {year} {1994})}\BibitemShut
  {NoStop}%
\bibitem [{\citenamefont {Rigol}\ \emph {et~al.}(2008)\citenamefont {Rigol},
  \citenamefont {Dunjko},\ and\ \citenamefont
  {Olshanii}}]{rigol2008thermalization}%
  \BibitemOpen
  \bibfield  {author} {\bibinfo {author} {\bibfnamefont {M.}~\bibnamefont
  {Rigol}}, \bibinfo {author} {\bibfnamefont {V.}~\bibnamefont {Dunjko}}, \
  and\ \bibinfo {author} {\bibfnamefont {M.}~\bibnamefont {Olshanii}},\
  }\href@noop {} {\bibfield  {journal} {\bibinfo  {journal} {Nature}\ }\textbf
  {\bibinfo {volume} {452}},\ \bibinfo {pages} {854} (\bibinfo {year}
  {2008})}\BibitemShut {NoStop}%
\bibitem [{\citenamefont {Cardy}(2014)}]{Cardy2014}%
  \BibitemOpen
  \bibfield  {author} {\bibinfo {author} {\bibfnamefont {J.}~\bibnamefont
  {Cardy}},\ }\href {\doibase 10.1103/PhysRevLett.112.220401} {\bibfield
  {journal} {\bibinfo  {journal} {Phys. Rev. Lett.}\ }\textbf {\bibinfo
  {volume} {112}},\ \bibinfo {pages} {220401} (\bibinfo {year}
  {2014})}\BibitemShut {NoStop}%
\bibitem [{\citenamefont {Garrison}\ and\ \citenamefont
  {Grover}(2018)}]{Garrison2018}%
  \BibitemOpen
  \bibfield  {author} {\bibinfo {author} {\bibfnamefont {J.~R.}\ \bibnamefont
  {Garrison}}\ and\ \bibinfo {author} {\bibfnamefont {T.}~\bibnamefont
  {Grover}},\ }\href {\doibase 10.1103/PhysRevX.8.021026} {\bibfield  {journal}
  {\bibinfo  {journal} {Phys. Rev. X}\ }\textbf {\bibinfo {volume} {8}},\
  \bibinfo {pages} {021026} (\bibinfo {year} {2018})}\BibitemShut {NoStop}%
\bibitem [{\citenamefont {Rigol}\ and\ \citenamefont
  {Srednicki}(2012)}]{Rigol-2012}%
  \BibitemOpen
  \bibfield  {author} {\bibinfo {author} {\bibfnamefont {M.}~\bibnamefont
  {Rigol}}\ and\ \bibinfo {author} {\bibfnamefont {M.}~\bibnamefont
  {Srednicki}},\ }\href {\doibase 10.1103/PhysRevLett.108.110601} {\bibfield
  {journal} {\bibinfo  {journal} {Phys. Rev. Lett.}\ }\textbf {\bibinfo
  {volume} {108}},\ \bibinfo {pages} {110601} (\bibinfo {year}
  {2012})}\BibitemShut {NoStop}%
\bibitem [{\citenamefont {Kaufman}\ \emph {et~al.}(2016)\citenamefont
  {Kaufman}, \citenamefont {Tai}, \citenamefont {Lukin}, \citenamefont
  {Rispoli}, \citenamefont {Schittko}, \citenamefont {Preiss},\ and\
  \citenamefont {Greiner}}]{Kaufman2016}%
  \BibitemOpen
  \bibfield  {author} {\bibinfo {author} {\bibfnamefont {A.~M.}\ \bibnamefont
  {Kaufman}}, \bibinfo {author} {\bibfnamefont {M.~E.}\ \bibnamefont {Tai}},
  \bibinfo {author} {\bibfnamefont {A.}~\bibnamefont {Lukin}}, \bibinfo
  {author} {\bibfnamefont {M.}~\bibnamefont {Rispoli}}, \bibinfo {author}
  {\bibfnamefont {R.}~\bibnamefont {Schittko}}, \bibinfo {author}
  {\bibfnamefont {P.~M.}\ \bibnamefont {Preiss}}, \ and\ \bibinfo {author}
  {\bibfnamefont {M.}~\bibnamefont {Greiner}},\ }\href {\doibase
  10.1126/science.aaf6725} {\bibfield  {journal} {\bibinfo  {journal}
  {Science}\ }\textbf {\bibinfo {volume} {353}},\ \bibinfo {pages} {794}
  (\bibinfo {year} {2016})},\ \Eprint
  {http://arxiv.org/abs/http://science.sciencemag.org/content/353/6301/794.full.pdf}
  {http://science.sciencemag.org/content/353/6301/794.full.pdf} \BibitemShut
  {NoStop}%
\bibitem [{\citenamefont {von Keyserlingk}\ \emph {et~al.}(2018)\citenamefont
  {von Keyserlingk}, \citenamefont {Rakovszky}, \citenamefont {Pollmann},\ and\
  \citenamefont {Sondhi}}]{Keyserlingk2017}%
  \BibitemOpen
  \bibfield  {author} {\bibinfo {author} {\bibfnamefont {C.~W.}\ \bibnamefont
  {von Keyserlingk}}, \bibinfo {author} {\bibfnamefont {T.}~\bibnamefont
  {Rakovszky}}, \bibinfo {author} {\bibfnamefont {F.}~\bibnamefont {Pollmann}},
  \ and\ \bibinfo {author} {\bibfnamefont {S.~L.}\ \bibnamefont {Sondhi}},\
  }\href {\doibase 10.1103/PhysRevX.8.021013} {\bibfield  {journal} {\bibinfo
  {journal} {Phys. Rev. X}\ }\textbf {\bibinfo {volume} {8}},\ \bibinfo {pages}
  {021013} (\bibinfo {year} {2018})}\BibitemShut {NoStop}%
\bibitem [{\citenamefont {Nahum}\ \emph {et~al.}(2018)\citenamefont {Nahum},
  \citenamefont {Vijay},\ and\ \citenamefont {Haah}}]{Nahum2017}%
  \BibitemOpen
  \bibfield  {author} {\bibinfo {author} {\bibfnamefont {A.}~\bibnamefont
  {Nahum}}, \bibinfo {author} {\bibfnamefont {S.}~\bibnamefont {Vijay}}, \ and\
  \bibinfo {author} {\bibfnamefont {J.}~\bibnamefont {Haah}},\ }\href {\doibase
  10.1103/PhysRevX.8.021014} {\bibfield  {journal} {\bibinfo  {journal} {Phys.
  Rev. X}\ }\textbf {\bibinfo {volume} {8}},\ \bibinfo {pages} {021014}
  (\bibinfo {year} {2018})}\BibitemShut {NoStop}%
\bibitem [{\citenamefont {Nahum}\ \emph {et~al.}(2017)\citenamefont {Nahum},
  \citenamefont {Ruhman},\ and\ \citenamefont {Huse}}]{nrh}%
  \BibitemOpen
  \bibfield  {author} {\bibinfo {author} {\bibfnamefont {A.}~\bibnamefont
  {Nahum}}, \bibinfo {author} {\bibfnamefont {J.}~\bibnamefont {Ruhman}}, \
  and\ \bibinfo {author} {\bibfnamefont {D.~A.}\ \bibnamefont {Huse}},\
  }\href@noop {} {\bibfield  {journal} {\bibinfo  {journal} {arXiv preprint
  arXiv:1705.10364}\ } (\bibinfo {year} {2017})}\BibitemShut {NoStop}%
\bibitem [{\citenamefont {Khemani}\ \emph {et~al.}(2018)\citenamefont
  {Khemani}, \citenamefont {Vishwanath},\ and\ \citenamefont
  {Huse}}]{Khemani2017}%
  \BibitemOpen
  \bibfield  {author} {\bibinfo {author} {\bibfnamefont {V.}~\bibnamefont
  {Khemani}}, \bibinfo {author} {\bibfnamefont {A.}~\bibnamefont {Vishwanath}},
  \ and\ \bibinfo {author} {\bibfnamefont {D.~A.}\ \bibnamefont {Huse}},\
  }\href {\doibase 10.1103/PhysRevX.8.031057} {\bibfield  {journal} {\bibinfo
  {journal} {Phys. Rev. X}\ }\textbf {\bibinfo {volume} {8}},\ \bibinfo {pages}
  {031057} (\bibinfo {year} {2018})}\BibitemShut {NoStop}%
\bibitem [{\citenamefont {Rakovszky}\ \emph {et~al.}(2018)\citenamefont
  {Rakovszky}, \citenamefont {Pollmann},\ and\ \citenamefont {von
  Keyserlingk}}]{rpv}%
  \BibitemOpen
  \bibfield  {author} {\bibinfo {author} {\bibfnamefont {T.}~\bibnamefont
  {Rakovszky}}, \bibinfo {author} {\bibfnamefont {F.}~\bibnamefont {Pollmann}},
  \ and\ \bibinfo {author} {\bibfnamefont {C.~W.}\ \bibnamefont {von
  Keyserlingk}},\ }\href {\doibase 10.1103/PhysRevX.8.031058} {\bibfield
  {journal} {\bibinfo  {journal} {Phys. Rev. X}\ }\textbf {\bibinfo {volume}
  {8}},\ \bibinfo {pages} {031058} (\bibinfo {year} {2018})}\BibitemShut
  {NoStop}%
\bibitem [{\citenamefont {Pai}\ \emph {et~al.}(2018)\citenamefont {Pai},
  \citenamefont {Pretko},\ and\ \citenamefont {Nandkishore}}]{pai2018}%
  \BibitemOpen
  \bibfield  {author} {\bibinfo {author} {\bibfnamefont {S.}~\bibnamefont
  {Pai}}, \bibinfo {author} {\bibfnamefont {M.}~\bibnamefont {Pretko}}, \ and\
  \bibinfo {author} {\bibfnamefont {R.~M.}\ \bibnamefont {Nandkishore}},\
  }\href@noop {} {\bibfield  {journal} {\bibinfo  {journal} {arXiv preprint
  arXiv:1807.09776}\ } (\bibinfo {year} {2018})}\BibitemShut {NoStop}%
\bibitem [{\citenamefont {{Chan}}\ \emph {et~al.}(2017)\citenamefont {{Chan}},
  \citenamefont {{De Luca}},\ and\ \citenamefont {{Chalker}}}]{Chan2017}%
  \BibitemOpen
  \bibfield  {author} {\bibinfo {author} {\bibfnamefont {A.}~\bibnamefont
  {{Chan}}}, \bibinfo {author} {\bibfnamefont {A.}~\bibnamefont {{De Luca}}}, \
  and\ \bibinfo {author} {\bibfnamefont {J.~T.}\ \bibnamefont {{Chalker}}},\
  }\href@noop {} {\bibfield  {journal} {\bibinfo  {journal} {ArXiv e-prints}\ }
  (\bibinfo {year} {2017})},\ \Eprint {http://arxiv.org/abs/1712.06836}
  {arXiv:1712.06836 [cond-mat.stat-mech]} \BibitemShut {NoStop}%
\bibitem [{\citenamefont {Chen}\ and\ \citenamefont
  {Zhou}(2018)}]{chen2018quantum}%
  \BibitemOpen
  \bibfield  {author} {\bibinfo {author} {\bibfnamefont {X.}~\bibnamefont
  {Chen}}\ and\ \bibinfo {author} {\bibfnamefont {T.}~\bibnamefont {Zhou}},\
  }\href@noop {} {\bibfield  {journal} {\bibinfo  {journal} {arXiv preprint
  arXiv:1808.09812}\ } (\bibinfo {year} {2018})}\BibitemShut {NoStop}%
\bibitem [{\citenamefont {Bertini}\ \emph {et~al.}(2019)\citenamefont
  {Bertini}, \citenamefont {Kos},\ and\ \citenamefont {Prosen}}]{bertini}%
  \BibitemOpen
  \bibfield  {author} {\bibinfo {author} {\bibfnamefont {B.}~\bibnamefont
  {Bertini}}, \bibinfo {author} {\bibfnamefont {P.}~\bibnamefont {Kos}}, \ and\
  \bibinfo {author} {\bibfnamefont {T.~c.~v.}\ \bibnamefont {Prosen}},\ }\href
  {\doibase 10.1103/PhysRevX.9.021033} {\bibfield  {journal} {\bibinfo
  {journal} {Phys. Rev. X}\ }\textbf {\bibinfo {volume} {9}},\ \bibinfo {pages}
  {021033} (\bibinfo {year} {2019})}\BibitemShut {NoStop}%
\bibitem [{\citenamefont {{\v{Z}}nidari{\v{c}}}\ \emph
  {et~al.}(2012)\citenamefont {{\v{Z}}nidari{\v{c}}} \emph
  {et~al.}}]{znidaric2012}%
  \BibitemOpen
  \bibfield  {author} {\bibinfo {author} {\bibfnamefont {M.}~\bibnamefont
  {{\v{Z}}nidari{\v{c}}}} \emph {et~al.},\ }\href@noop {} {\bibfield  {journal}
  {\bibinfo  {journal} {Journal of Physics A: Mathematical and Theoretical}\
  }\textbf {\bibinfo {volume} {45}},\ \bibinfo {pages} {125204} (\bibinfo
  {year} {2012})}\BibitemShut {NoStop}%
\bibitem [{\citenamefont {Yang}\ \emph {et~al.}(2015)\citenamefont {Yang},
  \citenamefont {Chamon}, \citenamefont {Hamma},\ and\ \citenamefont
  {Mucciolo}}]{hamma1}%
  \BibitemOpen
  \bibfield  {author} {\bibinfo {author} {\bibfnamefont {Z.-C.}\ \bibnamefont
  {Yang}}, \bibinfo {author} {\bibfnamefont {C.}~\bibnamefont {Chamon}},
  \bibinfo {author} {\bibfnamefont {A.}~\bibnamefont {Hamma}}, \ and\ \bibinfo
  {author} {\bibfnamefont {E.~R.}\ \bibnamefont {Mucciolo}},\ }\href {\doibase
  10.1103/PhysRevLett.115.267206} {\bibfield  {journal} {\bibinfo  {journal}
  {Phys. Rev. Lett.}\ }\textbf {\bibinfo {volume} {115}},\ \bibinfo {pages}
  {267206} (\bibinfo {year} {2015})}\BibitemShut {NoStop}%
\bibitem [{\citenamefont {Chamon}\ \emph {et~al.}(2014)\citenamefont {Chamon},
  \citenamefont {Hamma},\ and\ \citenamefont {Mucciolo}}]{hamma2}%
  \BibitemOpen
  \bibfield  {author} {\bibinfo {author} {\bibfnamefont {C.}~\bibnamefont
  {Chamon}}, \bibinfo {author} {\bibfnamefont {A.}~\bibnamefont {Hamma}}, \
  and\ \bibinfo {author} {\bibfnamefont {E.~R.}\ \bibnamefont {Mucciolo}},\
  }\href {\doibase 10.1103/PhysRevLett.112.240501} {\bibfield  {journal}
  {\bibinfo  {journal} {Phys. Rev. Lett.}\ }\textbf {\bibinfo {volume} {112}},\
  \bibinfo {pages} {240501} (\bibinfo {year} {2014})}\BibitemShut {NoStop}%
\bibitem [{\citenamefont {Yang}\ \emph {et~al.}(2017)\citenamefont {Yang},
  \citenamefont {Hamma}, \citenamefont {Giampaolo}, \citenamefont {Mucciolo},\
  and\ \citenamefont {Chamon}}]{hamma3}%
  \BibitemOpen
  \bibfield  {author} {\bibinfo {author} {\bibfnamefont {Z.-C.}\ \bibnamefont
  {Yang}}, \bibinfo {author} {\bibfnamefont {A.}~\bibnamefont {Hamma}},
  \bibinfo {author} {\bibfnamefont {S.~M.}\ \bibnamefont {Giampaolo}}, \bibinfo
  {author} {\bibfnamefont {E.~R.}\ \bibnamefont {Mucciolo}}, \ and\ \bibinfo
  {author} {\bibfnamefont {C.}~\bibnamefont {Chamon}},\ }\href {\doibase
  10.1103/PhysRevB.96.020408} {\bibfield  {journal} {\bibinfo  {journal} {Phys.
  Rev. B}\ }\textbf {\bibinfo {volume} {96}},\ \bibinfo {pages} {020408}
  (\bibinfo {year} {2017})}\BibitemShut {NoStop}%
\bibitem [{\citenamefont {Mierzejewski}\ \emph {et~al.}(2013)\citenamefont
  {Mierzejewski}, \citenamefont {Prosen}, \citenamefont {Crivelli},\ and\
  \citenamefont {Prelov\ifmmode~\check{s}\else
  \v{s}\fi{}ek}}]{Mierzejewski2013}%
  \BibitemOpen
  \bibfield  {author} {\bibinfo {author} {\bibfnamefont {M.}~\bibnamefont
  {Mierzejewski}}, \bibinfo {author} {\bibfnamefont {T.}~\bibnamefont
  {Prosen}}, \bibinfo {author} {\bibfnamefont {D.}~\bibnamefont {Crivelli}}, \
  and\ \bibinfo {author} {\bibfnamefont {P.}~\bibnamefont
  {Prelov\ifmmode~\check{s}\else \v{s}\fi{}ek}},\ }\href {\doibase
  10.1103/PhysRevLett.110.200602} {\bibfield  {journal} {\bibinfo  {journal}
  {Phys. Rev. Lett.}\ }\textbf {\bibinfo {volume} {110}},\ \bibinfo {pages}
  {200602} (\bibinfo {year} {2013})}\BibitemShut {NoStop}%
\bibitem [{sup()}]{suppmat}%
  \BibitemOpen
  \href@noop {} {}\bibinfo {note} {See online supplemental material for
  details.}\BibitemShut {Stop}%
\bibitem [{\citenamefont {Lieb}\ and\ \citenamefont
  {Robinson}(1972)}]{Lieb-1972}%
  \BibitemOpen
  \bibfield  {author} {\bibinfo {author} {\bibfnamefont {E.~H.}\ \bibnamefont
  {Lieb}}\ and\ \bibinfo {author} {\bibfnamefont {D.~W.}\ \bibnamefont
  {Robinson}},\ }in\ \href@noop {} {\emph {\bibinfo {booktitle} {Statistical
  mechanics}}}\ (\bibinfo  {publisher} {Springer},\ \bibinfo {year} {1972})\
  pp.\ \bibinfo {pages} {425--431}\BibitemShut {NoStop}%
\bibitem [{\citenamefont {Chan}\ \emph
  {et~al.}(2018{\natexlab{a}})\citenamefont {Chan}, \citenamefont {De~Luca},\
  and\ \citenamefont {Chalker}}]{Chan2018PRL}%
  \BibitemOpen
  \bibfield  {author} {\bibinfo {author} {\bibfnamefont {A.}~\bibnamefont
  {Chan}}, \bibinfo {author} {\bibfnamefont {A.}~\bibnamefont {De~Luca}}, \
  and\ \bibinfo {author} {\bibfnamefont {J.~T.}\ \bibnamefont {Chalker}},\
  }\href {\doibase 10.1103/PhysRevLett.121.060601} {\bibfield  {journal}
  {\bibinfo  {journal} {Phys. Rev. Lett.}\ }\textbf {\bibinfo {volume} {121}},\
  \bibinfo {pages} {060601} (\bibinfo {year} {2018}{\natexlab{a}})}\BibitemShut
  {NoStop}%
\bibitem [{\citenamefont {Kim}\ and\ \citenamefont {Huse}(2013)}]{Kim-2013}%
  \BibitemOpen
  \bibfield  {author} {\bibinfo {author} {\bibfnamefont {H.}~\bibnamefont
  {Kim}}\ and\ \bibinfo {author} {\bibfnamefont {D.~A.}\ \bibnamefont {Huse}},\
  }\href {\doibase 10.1103/PhysRevLett.111.127205} {\bibfield  {journal}
  {\bibinfo  {journal} {Phys. Rev. Lett.}\ }\textbf {\bibinfo {volume} {111}},\
  \bibinfo {pages} {127205} (\bibinfo {year} {2013})}\BibitemShut {NoStop}%
\bibitem [{\citenamefont {Kim}\ \emph {et~al.}(2015)\citenamefont {Kim},
  \citenamefont {Ba\~nuls}, \citenamefont {Cirac}, \citenamefont {Hastings},\
  and\ \citenamefont {Huse}}]{Kim-2015}%
  \BibitemOpen
  \bibfield  {author} {\bibinfo {author} {\bibfnamefont {H.}~\bibnamefont
  {Kim}}, \bibinfo {author} {\bibfnamefont {M.~C.}\ \bibnamefont {Ba\~nuls}},
  \bibinfo {author} {\bibfnamefont {J.~I.}\ \bibnamefont {Cirac}}, \bibinfo
  {author} {\bibfnamefont {M.~B.}\ \bibnamefont {Hastings}}, \ and\ \bibinfo
  {author} {\bibfnamefont {D.~A.}\ \bibnamefont {Huse}},\ }\href {\doibase
  10.1103/PhysRevE.92.012128} {\bibfield  {journal} {\bibinfo  {journal} {Phys.
  Rev. E}\ }\textbf {\bibinfo {volume} {92}},\ \bibinfo {pages} {012128}
  (\bibinfo {year} {2015})}\BibitemShut {NoStop}%
\bibitem [{\citenamefont {Brenes}\ \emph {et~al.}(2019)\citenamefont {Brenes},
  \citenamefont {Varma}, \citenamefont {Scardicchio},\ and\ \citenamefont
  {Girotto}}]{brenes2019}%
  \BibitemOpen
  \bibfield  {author} {\bibinfo {author} {\bibfnamefont {M.}~\bibnamefont
  {Brenes}}, \bibinfo {author} {\bibfnamefont {V.~K.}\ \bibnamefont {Varma}},
  \bibinfo {author} {\bibfnamefont {A.}~\bibnamefont {Scardicchio}}, \ and\
  \bibinfo {author} {\bibfnamefont {I.}~\bibnamefont {Girotto}},\ }\href@noop
  {} {\bibfield  {journal} {\bibinfo  {journal} {Computer Physics
  Communications}\ }\textbf {\bibinfo {volume} {235}},\ \bibinfo {pages} {477}
  (\bibinfo {year} {2019})}\BibitemShut {NoStop}%
\bibitem [{Note1()}]{Note1}%
  \BibitemOpen
  \bibinfo {note} {Because very small eigenvalues of $\rho $ are contaminated
  by machine precision, we define the width as twice the distance from the
  median to the 25th percentile of the entanglement spectrum (i.e., median to
  75th percentile of the spectrum of the RDM). This matches the interquartile
  range when both can be reliably computed.}\BibitemShut {Stop}%
\bibitem [{\citenamefont {Oganesyan}\ and\ \citenamefont
  {Huse}(2007)}]{oganesyan_huse}%
  \BibitemOpen
  \bibfield  {author} {\bibinfo {author} {\bibfnamefont {V.}~\bibnamefont
  {Oganesyan}}\ and\ \bibinfo {author} {\bibfnamefont {D.~A.}\ \bibnamefont
  {Huse}},\ }\href {\doibase 10.1103/PhysRevB.75.155111} {\bibfield  {journal}
  {\bibinfo  {journal} {Phys. Rev. B}\ }\textbf {\bibinfo {volume} {75}},\
  \bibinfo {pages} {155111} (\bibinfo {year} {2007})}\BibitemShut {NoStop}%
\bibitem [{\citenamefont {Guhr}\ \emph {et~al.}(1998)\citenamefont {Guhr},
  \citenamefont {Müller–Groeling},\ and\ \citenamefont
  {Weidenmüller}}]{GUHR1998}%
  \BibitemOpen
  \bibfield  {author} {\bibinfo {author} {\bibfnamefont {T.}~\bibnamefont
  {Guhr}}, \bibinfo {author} {\bibfnamefont {A.}~\bibnamefont
  {Müller–Groeling}}, \ and\ \bibinfo {author} {\bibfnamefont {H.~A.}\
  \bibnamefont {Weidenmüller}},\ }\href {\doibase
  https://doi.org/10.1016/S0370-1573(97)00088-4} {\bibfield  {journal}
  {\bibinfo  {journal} {Physics Reports}\ }\textbf {\bibinfo {volume} {299}},\
  \bibinfo {pages} {189 } (\bibinfo {year} {1998})}\BibitemShut {NoStop}%
\bibitem [{\citenamefont {Chen}\ and\ \citenamefont {Ludwig}(2018)}]{Chen2017}%
  \BibitemOpen
  \bibfield  {author} {\bibinfo {author} {\bibfnamefont {X.}~\bibnamefont
  {Chen}}\ and\ \bibinfo {author} {\bibfnamefont {A.~W.~W.}\ \bibnamefont
  {Ludwig}},\ }\href {\doibase 10.1103/PhysRevB.98.064309} {\bibfield
  {journal} {\bibinfo  {journal} {Phys. Rev. B}\ }\textbf {\bibinfo {volume}
  {98}},\ \bibinfo {pages} {064309} (\bibinfo {year} {2018})}\BibitemShut
  {NoStop}%
\bibitem [{\citenamefont {Atas}\ \emph {et~al.}(2013)\citenamefont {Atas},
  \citenamefont {Bogomolny}, \citenamefont {Giraud},\ and\ \citenamefont
  {Roux}}]{Atas2013}%
  \BibitemOpen
  \bibfield  {author} {\bibinfo {author} {\bibfnamefont {Y.~Y.}\ \bibnamefont
  {Atas}}, \bibinfo {author} {\bibfnamefont {E.}~\bibnamefont {Bogomolny}},
  \bibinfo {author} {\bibfnamefont {O.}~\bibnamefont {Giraud}}, \ and\ \bibinfo
  {author} {\bibfnamefont {G.}~\bibnamefont {Roux}},\ }\href {\doibase
  10.1103/PhysRevLett.110.084101} {\bibfield  {journal} {\bibinfo  {journal}
  {Phys. Rev. Lett.}\ }\textbf {\bibinfo {volume} {110}},\ \bibinfo {pages}
  {084101} (\bibinfo {year} {2013})}\BibitemShut {NoStop}%
\bibitem [{Note2()}]{Note2}%
  \BibitemOpen
  \bibinfo {note} {F. Pollmann, private communication.}\BibitemShut {Stop}%
\bibitem [{\citenamefont {Ho}\ and\ \citenamefont {Abanin}(2017)}]{ho2017}%
  \BibitemOpen
  \bibfield  {author} {\bibinfo {author} {\bibfnamefont {W.~W.}\ \bibnamefont
  {Ho}}\ and\ \bibinfo {author} {\bibfnamefont {D.~A.}\ \bibnamefont
  {Abanin}},\ }\href {\doibase 10.1103/PhysRevB.95.094302} {\bibfield
  {journal} {\bibinfo  {journal} {Phys. Rev. B}\ }\textbf {\bibinfo {volume}
  {95}},\ \bibinfo {pages} {094302} (\bibinfo {year} {2017})}\BibitemShut
  {NoStop}%
\bibitem [{\citenamefont {Rowlands}\ and\ \citenamefont
  {Lamacraft}(2018)}]{lamacraft2018}%
  \BibitemOpen
  \bibfield  {author} {\bibinfo {author} {\bibfnamefont {D.~A.}\ \bibnamefont
  {Rowlands}}\ and\ \bibinfo {author} {\bibfnamefont {A.}~\bibnamefont
  {Lamacraft}},\ }\href@noop {} {\bibfield  {journal} {\bibinfo  {journal}
  {arXiv preprint arXiv:1806.01723}\ } (\bibinfo {year} {2018})}\BibitemShut
  {NoStop}%
\bibitem [{\citenamefont {Knap}(2018)}]{knap2018}%
  \BibitemOpen
  \bibfield  {author} {\bibinfo {author} {\bibfnamefont {M.}~\bibnamefont
  {Knap}},\ }\href@noop {} {\bibfield  {journal} {\bibinfo  {journal} {arXiv
  preprint arXiv:1806.04686}\ } (\bibinfo {year} {2018})}\BibitemShut {NoStop}%
\bibitem [{\citenamefont {Xu}\ and\ \citenamefont {Swingle}(2018)}]{xu2018a}%
  \BibitemOpen
  \bibfield  {author} {\bibinfo {author} {\bibfnamefont {S.}~\bibnamefont
  {Xu}}\ and\ \bibinfo {author} {\bibfnamefont {B.}~\bibnamefont {Swingle}},\
  }\href@noop {} {\bibfield  {journal} {\bibinfo  {journal} {arXiv preprint
  arXiv:1802.00801}\ } (\bibinfo {year} {2018})}\BibitemShut {NoStop}%
\bibitem [{\citenamefont {Rakovzsky}\ \emph {et~al.}()\citenamefont
  {Rakovzsky}, \citenamefont {Gopalakrishnan}, \citenamefont {Parameswaran},\
  and\ \citenamefont {Pollmann}}]{rgpp}%
  \BibitemOpen
  \bibfield  {author} {\bibinfo {author} {\bibfnamefont {T.}~\bibnamefont
  {Rakovzsky}}, \bibinfo {author} {\bibfnamefont {S.}~\bibnamefont
  {Gopalakrishnan}}, \bibinfo {author} {\bibfnamefont {S.~A.}\ \bibnamefont
  {Parameswaran}}, \ and\ \bibinfo {author} {\bibfnamefont {F.}~\bibnamefont
  {Pollmann}},\ }\href@noop {} {}\bibinfo {note} {Unpublished}\BibitemShut
  {NoStop}%
\bibitem [{\citenamefont {Chan}\ \emph
  {et~al.}(2018{\natexlab{b}})\citenamefont {Chan}, \citenamefont {De~Luca},\
  and\ \citenamefont {Chalker}}]{cdc_prl}%
  \BibitemOpen
  \bibfield  {author} {\bibinfo {author} {\bibfnamefont {A.}~\bibnamefont
  {Chan}}, \bibinfo {author} {\bibfnamefont {A.}~\bibnamefont {De~Luca}}, \
  and\ \bibinfo {author} {\bibfnamefont {J.~T.}\ \bibnamefont {Chalker}},\
  }\href {\doibase 10.1103/PhysRevLett.121.060601} {\bibfield  {journal}
  {\bibinfo  {journal} {Phys. Rev. Lett.}\ }\textbf {\bibinfo {volume} {121}},\
  \bibinfo {pages} {060601} (\bibinfo {year} {2018}{\natexlab{b}})}\BibitemShut
  {NoStop}%
\end{thebibliography}%

\end{document}